\documentclass[review]{elsarticle}
\usepackage{amsmath,amssymb}
\usepackage{float}
\usepackage{array}
\usepackage[numbers]{natbib}
\setcitestyle{sort, square, comma}
\usepackage{graphicx}
\usepackage{caption}
\usepackage{lineno,hyperref}
\usepackage{amsmath}
\usepackage{graphicx}
\usepackage{xcolor}
\usepackage{epstopdf}
\usepackage[font=small]{caption}
\usepackage[labelformat=simple,font=footnotesize]{subcaption}

\modulolinenumbers[5]

\journal{Journal of Computational Physics}

\begin{document}

\begin{frontmatter}



%
%

%

\title{A simple cure for numerical shock instability in HLLC Riemann solver}
\author[mymainaddress]{Sangeeth Simon}
\author[mymainaddress]{Mandal J.C. \corref{mycorrespondingauthor}}
\address[mymainaddress]{Department of Aerospace Engineering, Indian Institute of Technology Bombay, Mumbai, India-400076}

 \cortext[mycorrespondingauthor]{Corresponding author}
\ead{mandal@iitb.ac.in}

\begin{abstract}
The Harten-Lax-van Leer with contact (HLLC) scheme is known to be plagued by various forms of numerical shock instabilities. 
In this paper, we propose a new framework for developing shock stable, contact and shear preserving approximate Riemann solvers based on the HLLC scheme for the Euler system of equations.
The proposed framework termed as HLLC-SWM (\textbf{S}elective \textbf{W}ave \textbf{M}odified) scheme identifies and increases the magnitude of the inherent diffusive HLL component within the HLLC scheme in the vicinity of a shock wave while leaving its antidiffusive component unmodified to retain accuracy on linearly degenerate wavefields. We present two strategies to compute the requisite supplementary dissipation which results in HLLC-SWM-E 
and HLLC-SWM-P variants.  
Through a linear perturbation analysis of the HLLC-SWM framework, we clarify how 
the additional dissipation introduced in this way helps in damping of unphysical perturbations in primitive quantities under a derived CFL constraint.
A matrix based stability analysis of a steady two-dimensional normal shock is used to show that both variants of the HLLC-SWM scheme are shock stable over a 
wide range of inlet Mach numbers. Results from standard test cases demonstrate that the HLLC-SWM schemes are capable of computing shock stable solutions on a 
variety of problems while retaining positivity and exact inviscid contact ability. On viscous flows, while the HLLC-SWM-P variant is quite accurate, the 
HLLC-SWM-E variant introduces slight inaccuracy which can be corrected through a simple Mach number based switching function.
\end{abstract}


\end{frontmatter}


\section{Introduction}

Accurate and robust computation of high-speed gas dynamical flows continues to be a challenge due to the presence of strong shocks, expansion fans, contact surfaces and their mutual interactions. 
The approximate Riemann solvers are one of the most popular methods available for numerical computations of such flows. However, it is well known that the contact and shear 
preserving variants of these schemes like the Roe \cite{roe1981}, HLLEM \cite{einfeldt1991} and HLLC \cite{toro1994} are susceptible to numerical shock instability during
multidimensional computations while the linear wavefield dissipative schemes like the HLL \cite{hll1983} and the HLLE \cite{einfeldt1988} are free from them.
These essentially multidimensional phenomena are primarily observed during the simulation of strong
grid aligned normal shocks. Benign perturbations in the initial conditions are rapidly amplified in the numerical shock region leading to spurious
solutions. Most commonly, these instabilities manifest as oscillations in the shock profile, polluted aftershock values, growth in error norms in case of steady state problems and in extreme cases complete breakdown of the solution. 
Additionally, the ocurrences of these instabilities are also
highly sensitive to parameters like grid aspect ratio \cite{henderson2007}, inflow Mach number\cite{dumbser2004}, 
numerical shock structure \cite{dumbser2004,chauvat2005}, perturbations in upstream supersonic flow \cite{dumbser2004} and order of accuracy of the solution \cite{gressier2000}
among other things.  A detailed catalouge of these 
failures can be found in  \cite{dumbser2004,quirk1994,pandolfi2001}.

Quirk \cite{quirk1994} conjectured that instability is caused because of coupling between pressure perturbations and density perturbations along the length of the shock which affects the numerical shock speed.  Dumbser et al \cite{dumbser2004} developed a useful matrix based stability analysis that can predict shock instability for a 
given scheme under various grid configurations, inlet Mach numbers and boundary conditions. 
Kun Xu \cite{xu2001} and others \cite{dumbser2004,chauvat2005,xie2017} located the source of instability to be within the numerical shock region. 
Some researchers \cite{xie2017,farhad2006,shen2016,ren2003} identify that inadequate dissipation of shear wave could result in these instabilities.
Sanders et al \cite{sanders1998} analyzed the numerical truncation error of upwind schemes to establish that shock instabilities in
these schemes arise due to lack of cross-flow dissipation provided by these schemes in the vicinity of a 
shock wave. Shen et al \cite{shen2014} showed how the Roe scheme suffers an overall loss of numerical dissipation along a numerically perturbed shock front
due to the vanishing of terms associated with the contact and shear waves. Further, they demonstrated that the addition of supplementary
dissipation on interfaces normal to the shock front mitigated this problem and produced shock stable solutions while ensuring consistency in mass flux across the shock. 

Based on these observations, various methods have been proposed in the literature to cure approximate Riemann solvers of this deficiency.
All of these cures are based on introducing a certain amount of
numerical dissipation that successfully attenuates the instabilities while inadvertently smearing the contact and shear waves. Thus there are hybrid Riemann solvers like Roe-HLLE \cite{wu2010,dongfang2016,nishikawa2008} and
HLLC-HLL \cite{shen2016,huang2011,kim2009,kim2010} and rotated Riemann solvers \cite{ren2003,nishikawa2008,zhang2016} that treat these instabilities with some success.
While the hybrid methods require explicit switching sensors and careful choice of complementary Riemann solvers, the rotated methods are computationally at least twice as
expensive to implement as their grid aligned counterparts. As yet, one of the simplest methods to deal with shock instabilities in Roe scheme involves increasing its overall numerical dissipation by modifying the eigenvalues of its linearized Jacobian matrix; including the ones corresponding to the linear waves. 
Peery and Imlay \cite{peery1988} demonstrated a cure for 'Carbuncle phenomenon' in Roe scheme based on this idea. The modification they proposed used a direction based pressure gradient to tune the magnitude of the dissipation introduced. Lin \cite{lin1995} pointed out the limitations of this 
pressure sensor and proposed an improved version that works even on viscous flows.
Sanders et al \cite{sanders1998} suggested using the idea of an entropy fix on linear waves in conjunction with a simple \textquoteleft H\textquoteright shaped multidimensional stencil to estimate the requisite dissipation to cure the Roe scheme. Later Pandolfi et al \cite{pandolfi2001} pointed out a 
more efficient way to implement the same that avoids unwanted dissipation at shear layers for more anisotropic grids. However, as noted by some authors 
\cite{dumbser2004,gressier2000}, there is no physical justification for modifying the linear eigenvalues using an entropy fix.

Incidentally, a majority of cures for shock instability have been developed for the Roe scheme. However, Roe scheme suffers several drawbacks like requirement of an 
additional entropy fix on its nonlinear wave speeds to admit entropy satisfying solutions \cite{harten_hyman1983}, knowledge of the full eigenstructure of the 
flux Jacobians making it difficult to be extended to general
system of governing equations with complex equations of state \cite{balsara2010} and lack of positivity \cite{einfeldt1991}.
These shortcomings of Roe scheme 
has brought forth an increasing interest in another equally accurate and efficient contact and shear preserving Riemann solver - the HLLC Riemann solver \cite{toro1994}
which also, unfortunately, succumbs to various numerical shock instabilities. Hence the need for delivering the HLLC solver from these instabilities is paramount.

In this paper, we present a very simple and inexpensive strategy to develop a shock stable HLLC scheme. 
The strategy is termed HLLC-SWM (\textbf{S}elective \textbf{W}ave \textbf{M}odified). The novelty of the formulation lies in first decomposing the HLLC scheme into its diffusive HLL component and the antidiffusive component that restores the accuracy on linear waves. Once distinguished, we propose a strategy to adaptively increase  
the numerical dissipation associated with this inherent HLL component alone in the vicinity of shocks through appropriate modification of selected nonlinear wave speeds.
In contrast to the cures proposed in \cite{pandolfi2001,sanders1998,lin1995}, the present method only targets nonlinear wave speeds that resides in the diffusive component while leaving the middle wave speed which is crucial to the contact and shear preserving ability of the HLLC solver completely undisturbed. 
Various theoretical analyses and an array of numerical test problems confirm that this simple technique is found to provide sufficient dissipation in the vicinity of the shock to eliminate shock instabilities.

The outline of the paper is as follows. In Sections \ref{sec:governing_equations} and \ref{sec: FVM_framework} a brief outline of the governing equation and the Finite Volume framework used in the present work is discussed.  Section \ref{sec:reviewofHLLCandHLL} reviews the HLL and the HLLC Riemann solvers. In Section \ref{sec:formulation} 
we discuss the formulation of the HLLC-SWM framework and prove the effectiveness of the proposed strategy in increasing the overall dissipation of the HLLC scheme. 
In Section \ref{sec:quirkanalysis}, a linear analysis technique is used to investigate how the HLLC-SWM framework is effective in suppressing perturbations in primitive quantities. 
In Sections \ref{sec:estimationofepsilon} and \ref{sec:estimationofalpha}, we describe two methods to determine the amount of dissipation required by the 
HLLC-SWM formulation to suppress shock instabilities over a wide range of Mach numbers. In Section 
\ref{sec:numericaltests} the HLLC-SWM formulation is extensively tested on a series of numerical test problems. Some concluding remarks are given in Section 
\ref{sec:conclusions}.

\section{Governing equations}
\label{sec:governing_equations}
The governing equations for two dimensional inviscid compressible flow can be expressed in their conservative form as, 
\begin{align}
 \frac{\partial \mathbf{\acute{U}}}{\partial t} + \frac{\partial \mathbf{\acute{F}(U)}}{\partial x} + \frac{\partial \mathbf{\acute{G}(U)}}{\partial y} = 0
 \label{equ:EE-differentialform}
\end{align}
where $\mathbf{\acute{U}}, \mathbf{\acute{F}(U)} , \mathbf{\acute{G}(U)}$ are the vector of conserved variables and x and y directional fluxes respectively. These are 
given by, 

\begin{align}
 \mathbf{\acute{U}} = \left [ \begin{array}{c}
\rho \\
\rho u\\
\rho v\\
\rho E
\end{array}\right ],
\mathbf{\acute{F}(U)} = \left [ \begin{array}{c}
\rho u \\
\rho u^2 + p\\
\rho u v\\
(\rho E + p)u
\end{array}\right ],
\mathbf{\acute{G}(U)}= \left [ \begin{array}{c}
\rho v \\
\rho u v\\
\rho v^2 + p\\
(\rho E + p)v
\end{array}\right ]
\end{align}
In the above expressions, $\rho, u, v, p$ and $E$ stands for global density, x-velocity, y-velocity, pressure and specific total energy. The system of 
equations are closed through the equation of state,
\begin{align}
 p = (\gamma-1)\left(\rho E - \frac{1}{2}\rho(u^2 + v^2)\right)
\end{align}
where $\gamma$ is the ratio of specific heats. Present work assumes a calorifically perfect gas with $\gamma =1.4$. A particularly useful form of the 
Eq.(\ref{equ:EE-differentialform}) is the integral form given by,

\begin{align}
 \frac{\partial}{\partial t} \int_{\varOmega} \mathbf{\acute{U}} dxdy + \oint_{d\varOmega} [(\mathbf{\acute{F},\acute{G}}).\mathbf{n}] dl = 0 
 \label{eqn:EE-integralform}
\end{align}
where $\varOmega$ denotes a control volume over which Eq.(\ref{eqn:EE-integralform}) describes a Finite Volume balance of the conserved
quantities, $dx$ and $dy$
denotes the x and y dimensions of the control volume respectively, $d\varOmega$ denotes the boundary 
surface of the control volume and $dl$ denotes an infinitesimally small element on $d\varOmega$. $\mathbf{n}$ is the outward pointing unit normal vector to the surface $d\varOmega$. 

\section{Finite volume discretization}
\label{sec:FVM_framework}
\begin{figure}[H]
	    \centering
	    \includegraphics[scale=0.3]{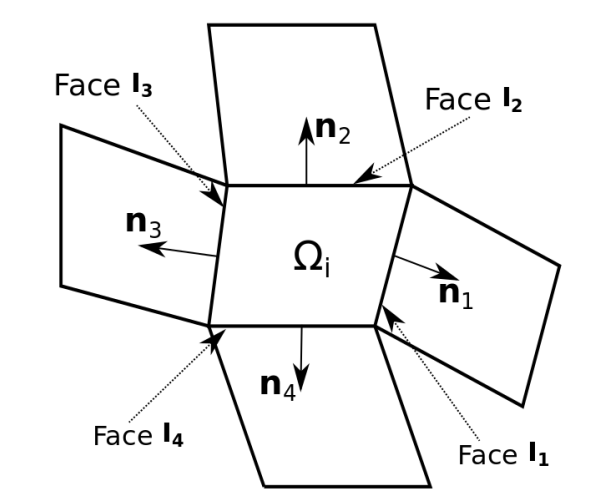}
	    \caption{Typical control volume $i$ with its associated interfaces $I_k$ and respective normal vectors $\mathbf{n}_k$.}
	    \label{fig:finitevolume}
	    \end{figure}

In this paper we seek a Finite Volume numerical solution of Eq.(\ref{eqn:EE-integralform}) by discretizing the equation on a computational 
mesh consisting of structured quadrilateral cells as shown in Fig.(\ref{fig:finitevolume}). For a typical cell $i$ belonging to this mesh, a 
semi-discretized version
of such a solution can be written as,

\begin{align}
 \frac{d\mathbf{U}_i}{dt} = -\frac{1}{|\varOmega|_i} \sum_{k=1}^{4} [({\mathbf{\acute{F},\acute{G}}})_k.\mathbf{n}_k] \Delta s_k
 \label{eqn:EE-partialdiscretized}
\end{align}
where ${\mathbf{U}_i}$ is an appropriate cell averaged conserved state vector, $({\mathbf{\acute{F},\acute{G}}})_k$ denotes the flux vector at the 
mid point of each interface $I_k$ while $\mathbf{n}_k$ and $\Delta s_k$ denotes the unit normal vector  and the length of each $I_k$ interface 
respectively. These are shown in Fig.(\ref{fig:finitevolume}). The interface flux $({\mathbf{\acute{F},\acute{G}}})_k.\mathbf{n}_k$ can be obtained by various methods. 
One of the most popular class of methods 
to compute this are the approximate Riemann solvers. A conventional two state approximate Riemann solver uses the 
rotational invariance property of Euler equations to express the term $({\mathbf{\acute{F},\acute{G}}})_k.\mathbf{n}_k$ as, 

\begin{align}
 \frac{d\mathbf{U}_i}{dt} = -\frac{1}{|\varOmega|_i} \sum_{k,m=1}^{4} [\mathbf{T}^{-1}_{k} \mathbf{F} (\mathbf{U_L},\mathbf{U_R})]\Delta s_k
\label{eqn:EE-rotationalinvariance}
 \end{align}
 where $ \mathbf{U_L} = \mathbf{T}_{k}(\mathbf{U}_i), \mathbf{U_R}=\mathbf{T}_{k}(\mathbf{U}_m)$ indicates the initial conditions of a local
Riemann problem across $k^{th}$ interface formed by the $i^{th}$ cell and its neighbouring cell $m$. The matrices $\mathbf{T}_{k}$ and $\mathbf{T}_{k}^{-1}$ 
are rotation matrices at the $k^{th}$ interface and is described in \cite{toro2009}.
 
%

In the next section we briefly describe two approximate Riemann solvers namely the HLL and the HLLC that can be used to estimate the flux 
$\mathbf{F} (\mathbf{U_L},\mathbf{U_R})$
at a given interface. To avoid extra notations, henceforth $\mathbf{F}$ simply represent the local Riemann flux at any interface with outward pointing normal $\mathbf{n}$.

\section{A review of two classic Riemann solvers}
\label{sec:reviewofHLLCandHLL}
The following subsections briefs two classic Riemann solvers on which the present work builds upon. 

\subsection{The HLL Riemann solver}
The HLL Riemann solver, devised by Harten, Lax and van Leer \cite{hll1983} assumes a wave structure consisting of two waves that
seperates three constant states. These two waves represent
the genuinely nonlinear waves like shocks and expansion fans. The average state $\mathbf{U}^{*}_{HLL}$ can be derived through application of integral form of 
conservation laws as, 

\begin{align}
  \mathbf{U}_{*}^{HLL}  &= \frac{S_R \mathbf{U}_R - S_L \mathbf{U}_L + \mathbf{F}_L - \mathbf{F}_R }{S_R - S_L}
  \label{eqn:HLLstate}
\end{align}
Using Rankine-Hugoniot conditions, the HLL Riemann flux can then be written as, 

\begin{align}
    \mathbf{F}_{HLL}=     \frac{S_R \mathbf{F}_L - S_L \mathbf{F}_R + S_LS_R (\mathbf{U}_R - \mathbf{U}_L) }{S_R - S_L}
\label{eqn:hllformualtion}
\end{align}
Here $S_L$ and $S_R$ are numerical approximations to the speeds of the left most and right most running characteristics that emerge as the solution of 
Riemann problem at the cell interface of interest.
It has been shown that under appropriate choice of wavespeeds $S_L$ and $S_R$, the HLL scheme is both positivity preserving and entropy
satisfying \cite{einfeldt1988}. In this work, $S_L$ and $S_R$ are computed as \cite{einfeldt1988},
\begin{align}
\nonumber
S_L = min(0,u_{nL}-a_L, \tilde{u}_n-\tilde{a})\\
 S_R = max(0,u_{nR}+a_R, \tilde{u}_n +\tilde{a})
 \label{eqn:HLLwavespeedestimate}
\end{align}
where $u_{nL,R}$ are the normal velocity across an interface, $a_{L,R}$ are the respective sonic speeds and $\tilde{u}_n,\tilde{a}$ are the standard Roe averaged 
quantities at the interface \cite{roe1981}. Using these wavespeed estimates, the HLL scheme is also known as the HLLE scheme \cite{einfeldt1988}.
It has been 
shown that under these wave speed, the HLLE scheme is both positivity preserving and entropy satisfying. 
Although the HLLE scheme is very robust for shock capturing applications, one of its major
drawback is its inability to resolve the linearly degenerate waves of the Euler system that govern the contact and the shear phenomenon. Such an inability severely restricts the scheme from being used 
in a full Navier-Stokes solver because linear wave resolution is generally regarded as a prerequisite for accurate resolution of viscous 
phenomenon \cite{vanleer1987,toro_vaz2012}. Gressier et al \cite{gressier2000} have 
conjectured that upwind schemes that posess exact contact ability cannot be free from shock instabilities. Hence by corollary, the shock stable behaviour of the HLLE scheme 
is attributed to its linear wave dissipative nature \cite{pandolfi2001}.

\subsection{The HLLC Riemann solver}
Toro et al \cite{toro1994} mitigated the short coming of the HLL Riemann solver by introducing the contact and shear preserving feature into it. The resulting solver is 
called the HLLC ( C for Contact) scheme. This improvement was achieved by adding a third 
wave to the one dimensional HLL wave structure, called a contact wave with speed $S_M$. Once again by using the integral form of conservation laws and appropriate
flow conditions relating to the middle wave, closed form expressions for the middle state quantities $\mathbf{U}_{*L/R}^{HLLC}$ can be derived. Based on this the HLLC 
interface flux $\mathbf{F}_{HLLC}$ can be written as, 

\begin{align}
    \mathbf{F}_{HLLC}= 
    \begin{cases}
    \mathbf{F}_L + S_L(\mathbf{U}_{*L}^{HLLC} - \mathbf{U}_L),& \text{if } S_L\leq 0 \leq S_M\\
    \mathbf{F}_R + S_R(\mathbf{U}_{*R}^{HLLC} - \mathbf{U}_R),& \text{if } S_M \leq 0 \leq S_R\\
    \end{cases}
\label{eqn:hllcformulation}
\end{align}
The two intermediate states $\mathbf{U}_{*L}^{HLLC}$ and $\mathbf{U}_{*R}^{HLLC}$ that exist across the additional wave $S_M$ are obtained as, 
\begin{align}
    \mathbf{U}_{*L/R}^{HLLC} &= \rho_{L/R} \left(\frac{S_{L/R}-u_{nL/R}}{S_{L/R}-S_M}\right) \left( \begin{array}{c}
                                                                              1\\
                                                                              S_M\\
                                                                              u_{tL/R}\\
                                                                              \frac{(\rho E)_{L/R}}{\rho_{L/R}} + (S_M-u_{nL/R})(S_M + \frac{p_{L/R}}{\rho_{L/R}(S_{L/R}-u_{nL/R})})                                                        
                                                                              \end{array} \right)
    \label{eqn:HLLCstate}
\end{align}
where $u_{tL/R}$ denote the local tangential velocity at an interface.	   
In the above expressions, $S_L$ and $S_R$ can be obtained using Eq.(\ref{eqn:HLLwavespeedestimate}). Batten et al \cite{batten1997} provides a
closed form expression for $S_M$ as,
\begin{align}
 \label{eqn:HLLCmiddlewaveestimate}
  S_M = \frac{p_R - p_L + \rho_Lu_L(S_L - u_L) - \rho_Ru_R(S_R - u_R)}{\rho_L(S_L - u_L) -\rho_R(S_R - u_R)}
\end{align}
The HLLC Riemann solver is one of the simplest known Riemann solver to be able to resolve all the waves of the Euler system accurately.
Like the HLL Riemann solver, it is also an entropy satisfying and positively conservative scheme under appropriate wave speed selection \cite{toro1994,batten1997}.
However, albeit all the attractive features the scheme
posesses, the HLLC Riemann solver is known to be highly susceptible to numerical shock instabilities. In the next section, a simple and inexpensive method of protecting 
the HLLC scheme from the shock instability phenomenon is presented. The idea involves first splitting the standard HLLC scheme into its diffusive HLL component  
and an antidiffusive 
component responsible for restoring linear wave accuracy to this HLL component. Then, a multidimensional solution dependent strategy is used to enhance the magnitude 
of this diffusive component in the vicinity of shocks through simple modifications of certain 
carefully chosen nonlinear wavespeed estimates. The resulting solver
is found to be completely free of numerical shock instabilities yet retaining the accuracy and efficiency of the original HLLC solver.

\section{Formulation of a shock stable HLLC scheme}
\label{sec:formulation}

We propose the following strategy to save the HLLC scheme from instabilities.  
We first decompose the HLLC scheme into its diffusive component and the antidiffusive component responsible for restoration of accuracy on contact and shear waves.
To achieve this we propose to rewrite the HLLC scheme as, 

\begin{align}
    \mathbf{F}_{HLLC}= 
\begin{cases}
    \mathbf{F}_{HLL} + S_L(\mathbf{U}_{*L}^{HLLC} - \mathbf{U}_{*}^{HLL}),& \text{if } S_L\leq 0 \leq S_M\\
    \mathbf{F}_{HLL} + S_R(\mathbf{U}_{*R}^{HLLC} - \mathbf{U}_{*}^{HLL}),& \text{if } S_M \leq 0 \leq S_R\\
\end{cases}
\label{eqn:hllc-modifiedform}
\end{align}
where $\mathbf{F}_{HLL} =  \mathbf{F}_{L/R} + S_{L/R} (\mathbf{U}^*_{HLL}- \mathbf{U}_{L/R})$. 
It is seen that $\mathbf{F}_{HLL}$ can be further recast into a second order central plus upwind dissipation form as, 
\begin{align}
   \mathbf{F}_{HLL} = \frac{1}{2}(\mathbf{F}_L + \mathbf{F}_R) + \mathbf{D}_{HLL}
   \label{eqn:HLL_centralplusdissipation}
\end{align}
where the numerical dissipation of the embedded HLL scheme is \cite{mandal2012},
\begin{align}
 \mathbf{D}_{HLL} = \frac{S_R+S_L}{2(S_R-S_L)}(\mathbf{F}_L-\mathbf{F}_R) - \frac{S_LS_R}{(S_R-S_L)}(\mathbf{U}_L-\mathbf{U}_R)
 \label{eqn:HLLdissipation}
\end{align}
This can also be written in an equivalent form inspired from \cite{schmidtmann2017} as,
\begin{align}
\mathbf{D}_{HLL} = \frac{|S_R|-|S_L|}{2(S_R-S_L)}(\mathbf{F}_L-\mathbf{F}_R) +  \frac{|S_L|S_R-|S_R|S_L}{2(S_R-S_L)}(\mathbf{U}_L-\mathbf{U}_R)
\label{eqn:HLLdissipation-absoluteform}
\end{align}
Rewriting HLLC scheme in the form presented in Eq.(\ref{eqn:hllc-modifiedform}) helps in clear identification of the diffusive and antidiffusive components 
that make up the scheme. While the overall disspation 
vector $\mathbf{D}_{HLL}$ occurs as a part of the HLL component of the flux, the accuracy on contact and shear waves that makes HLLC more accurate than the 
HLL solver is provided through the antidissipation vectors $S_{L/R}(\mathbf{U}^{*HLLC}_{L/R} - \mathbf{U}^{*HLL})$ in each case. 
Once the inherent dissipation vector $\mathbf{D}_{HLL}$ is identified, a strategy to increase the magnitude of it is sought near shock waves.
Davis \cite{davis1988} has noted that choice of the wavespeeds $S_L$ and $S_R$ can affect the amount of dissipation that is introduced into the 
HLL scheme. Hence it is envisioned that additional dissipation to deal with shock instabilities can be introduced into the HLLC scheme through
careful manipulations of the wavespeeds $S_L$ and $S_R$ present in $\mathbf{D}_{HLL}$.
In the present work, a modified dissipation vector $\overline{\mathbf{D}}_{HLL}$ is proposed, 
\begin{align}
\overline{\mathbf{D}}_{HLL} = \frac{|\overline{S_R}|-|\overline{S_L}|}{2(S_R-S_L)}(\mathbf{F}_L-\mathbf{F}_R) +  \frac{|\overline{S_L}|S_R-|\overline{S_R}|S_L}{2(S_R-S_L)}(\mathbf{U}_L-\mathbf{U}_R)
\label{eqn:HLLCSWMdissipation-absoluteform}
\end{align}
where, 
\begin{align}
\nonumber
 \overline{S_L} = S_L - \alpha\epsilon^{L}\\
 \overline{S_R} = S_R + \alpha\epsilon^{R}
 \label{eqn:hllc_modifiedwavespeedestimates}
\end{align}
$S_L$ and $S_R$ in the above equations are obtained from Eq.(\ref{eqn:HLLwavespeedestimate}). The term $\epsilon^{L,R}$ is an appropriate choice of dissipation 
of the order $\mathcal{O} (S_L,S_R)$ and $\alpha$ is a coefficient through which the 
quantity of $\epsilon^{L,R}$ introduced into $S_L, S_R$ can be controlled. Both $\epsilon^{L,R}$ and $\alpha$ are by definition nonnegative quantities and will be
defined completely later. It can be seen intutitvely that modification of wavespeeds in Eq.(\ref{eqn:hllc_modifiedwavespeedestimates}) 
amounts to spreading of the wave structure of the inherent HLL scheme thereby increasing the magnitude of $\mathbf{D}_{HLL}$ which will reflect as an increase in the
overall dissipation of the HLLC scheme.
Eqs.(\ref{eqn:hllc-modifiedform},\ref{eqn:HLL_centralplusdissipation},\ref{eqn:HLLCSWMdissipation-absoluteform} and \ref{eqn:hllc_modifiedwavespeedestimates})
describe the proposed modified HLLC scheme. 

Note that the contact wave speed $S_M$ and the wavespeeds $S_L, S_R$ associated with the antidiffusion component
are left completely unaffected by this modification. This is done purposefully to avoid affecting the contact and shear capturing ability of the HLLC solver.
It is important to emphasize here that although the technique proposed here aligns with the idea of wave speed modification to achieve shock stability as recommended in 
\cite{sanders1998}, it differs from it in that instead of targeting eigenvalues of the linear fields, the dissipation is added through the wave speed estimates of  
the nonlinear fields. This can be justified by the fact that in the two wave formulation of the HLL scheme, $S_L$ and $S_R$ are the numerical characteristics associated with the 
shock waves and hence any dissipation infused through them would directly affect shock capturing ability only.
Owing to this \textit{selective wave modification} strategy the proposed modified version of the HLLC scheme will be henceforth refered to in this work as 
HLLC-SWM (\textbf{S}elective \textbf{W}ave \textbf{M}odified) scheme.  
It is emphasized that the modification presented in 
Eq.(\ref{eqn:hllc_modifiedwavespeedestimates}) must be used in conjuction with the form of dissipation vector $\overline{\mathbf{D}}_{HLL}$ presented in 
Eq.(\ref{eqn:HLLCSWMdissipation-absoluteform}) for it to be effective. In the following subsection, we provide an analytical proof showing how the modification proposed in 
Eq.(\ref{eqn:hllc_modifiedwavespeedestimates}) manages to improve the overall dissipation of the HLLC scheme.

\subsection{Effect of $\epsilon^{L,R}$ on numerical dissipation of the HLLC scheme}
\label{sec:effectofepsilon_derivation}
In this section we aim to analytically demonstrate how the $\epsilon^{L,R}$ parameter introduced through modified wavespeeds in Eq.(\ref{eqn:hllc_modifiedwavespeedestimates})
into the dissipation vector form in Eq.(\ref{eqn:HLLCSWMdissipation-absoluteform}) amounts to overall increase in dissipation of the HLLC scheme.
Although the explanation is provided through an expression of $\mathbf{F}_{HLLC}$ corresponding to condition $ S_L\leq 0 \leq S_M$, it remains symmetrically valid 
for the case when $S_M \leq 0 \leq S_R$. As observed previously from Eq.(\ref{eqn:hllc-modifiedform}), the HLLC flux can be interpreted as a combination of a 
diffusive 
HLL flux plus an antidiffusive component responsible for accuracy on contact and shear waves as,

\begin{align}
\nonumber
 \mathbf{F}_{HLLC} = \underbrace{\mathbf{F}_L + S_L (\mathbf{U}^*_{HLL}- \mathbf{U}_L)}_{HLL\ Flux} + \underbrace{S_L(\mathbf{U}^*_{HLLC} - \mathbf{U}^*_{HLL})}_{HLLC\ Antidissipation}  
\end{align}
The numerical dissipation vector $\mathbf{D}_{HLL}$ of the this inherent HLL component can be expressed as in Eq.(\ref{eqn:HLLdissipation-absoluteform}),
\begin{align}
\mathbf{D}_{HLL} = \frac{|S_R|-|S_L|}{2(S_R-S_L)}(\mathbf{F}_L-\mathbf{F}_R) +  \frac{|S_L|S_R-|S_R|S_L}{2(S_R-S_L)}(\mathbf{U}_L-\mathbf{U}_R)
\label{eqn:HLLdissipation-absoluteform}
\end{align}
Let $a_0$ and $a_1$ be coefficients of $\mathbf{D}_{HLL}$ defined as,
\begin{align}
\nonumber
a_0 &= \frac{|S_R|-|S_L|}{2(S_R-S_L)} \\
a_1 &= \frac{|S_L|S_R - |S_R|S_L}{2(S_R-S_L)}
\label{eqn:dissipationcoefficients_original}
\end{align}
Let $\overline{a_0}$ and $\overline{a_1}$ be the corresponding coefficients of the modified dissipation vector 
$\overline{\mathbf{D}}_{HLL}$ given in Eq.(\ref{eqn:HLLCSWMdissipation-absoluteform})  
\begin{align}
\nonumber
\overline{a_0} &= \frac{|S_R+\epsilon^{R}|-|S_L-\epsilon^{L}|}{2(S_R-S_L)} \\
\overline{a_1} &= \frac{|S_L-\epsilon^{L}|S_R - |S_R+\epsilon^{R}|S_L}{2(S_R-S_L)}
\label{eqn:dissipationcoefficients_modified}
\end{align}
For the sake of convenience, in the above expressions, the value of $\alpha$ is taken to be 1.
Since in subsonic flows, $S_L<0$ and $S_R>0$, it becomes immediately evident that the denominator of $\overline{a_0}$ and $\overline{a_1}$ 
will be always postive. Further, coefficients in Eq.(\ref{eqn:dissipationcoefficients_modified}) can be simplified as,



\begin{align}
\nonumber
 \overline{a_0} &= \frac{|S_R|-|S_L|}{2(S_R-S_L)}+\frac{\epsilon^R - \epsilon^L}{2(S_R-S_L)} \\
 \overline{a_1} &= \frac{|S_L|S_R - |S_R|S_L}{2(S_R-S_L)} + \frac{\epsilon^LS_R-\epsilon^RS_L}{2(S_R-S_L)}
 \label{eqn:dissipationcoefficients_simplified}
\end{align}
Some interesting observations are forthcoming at this point. It is clear from Eq.(\ref{eqn:dissipationcoefficients_simplified}) that both coefficients 
$\overline{a_0}$ and $\overline{a_1}$ turn out to be respectively $a_0$ and $a_1$ with additional terms that are functions of the dissipation parameter $\epsilon^{L,R}$. 
If we choose $\epsilon^R>\epsilon^L$
then these terms are positive additions and we can ensure that $\overline{a_0}>a_0$ and $\overline{a_1}>a_1$. Note that such a constraint on $\epsilon^{L,R}$
can be imposed simply by setting say, $\epsilon^L \sim |S_L|$ and $\epsilon^R \sim |S_R|$ in a subsonic flow. 
Consider a simple choice $\epsilon^L=\epsilon^R=\epsilon$. Then the additional term in $\overline{a_0}$ vanish while the additional term in $\overline{a_1}$ becomes $\epsilon/2$. Thus for such a choice, it is seen that the
modification presented in Eq.(\ref{eqn:hllc_modifiedwavespeedestimates}) effectively affects only the $\mathbf{U}_L-\mathbf{U}_R$ vector (ie coefficient $a_1$) and nowhere else 
in the entire HLLC formulation. This fact can be exploited to reduce the cost associated with implementation of the present cure to an already existing HLLC Riemann 
solver code. 
The above discussion clearly indicates that the modified HLL dissipation vector $\overline{\mathbf{D}}_{HLL}$ provides an excellent framework 
to introduce adequate dissipation into the original HLLC Riemann solver. However as seen above, the definition of $\epsilon^{L,R}$ has a significant role to play in the effectiveness
of $\overline{\mathbf{D}}_{HLL}$.

In the next section we perform a linear perturbation analysis on the HLLC-SWM framework to reveal the intricate role of $\epsilon^{L,R}$ 
parameter in damping the numerical perturbations in flow field values which are known to be the prime instigators of these instabilities.

\section{Linear perturbation analysis of HLLC-SWM scheme}
\label{sec:quirkanalysis}
A technique of predicting whether a numerical flux function will produce shock unstable solution based on how the flux function evolves an initially perturbed
primitive variable profile was established by Quirk \cite{quirk1994} and others \cite{gressier2000,pandolfi2001}.
Quirk conjectured that \textit{schemes in which pressure perturbations feed into density perturbations are prone to produce shock instabilities}.  
In this regard, it is interesting to use this technique to study the evolution of primitive variable perturbations using the HLLC-SWM scheme. 
Of specific interest is in discerning the effect of the dissipation parameters $\epsilon^{L,R}$ and $\alpha$ on this evolution.
A steady mean flow with normalised 
state values of $\rho_0=1$, $u_0\neq0$, $v_0=0$ and $p_0=1$ is chosen as the base flow. 
On a typical y-directional stencil the cells are marked as \textquotedblleft even \textquotedblright and \textquotedblleft odd \textquotedblright \cite{pandolfi2001}.
The perturbations are introduced as a saw tooth profile in density, 
x-velocity and pressure variables. For a typical cell 'j' belonging to this stencil, these are initialized as,
\begin{align}
\begin{cases}
    \rho_j=\rho_0+\hat{\rho}, \;\;p_j=p_0+\hat{p},\;\;u_j=u_0+\hat{u},\;\;v_j=0 ,& \text{if } j\;\; is\;\; even\\
    \rho_j=\rho_0-\hat{\rho}, \;\;p_j=p_0-\hat{p},\;\;u_j=u_0-\hat{u},\;\;v_j=0 ,& \text{if } j\;\; is\;\; odd
\end{cases}
\label{eqn:sawtoothinitialconditions}
\end{align}
Such saw tooth profiles are characteristic of most shock unstable solutions. The crux of the analysis is to develop equations that describe the 
temporal evolution of these saw tooth perturbations described in Eq.(\ref{eqn:sawtoothinitialconditions}).  
Note that we avoid introducing perturbations $\hat{v}$ in y-directional velocity in the present analysis
because it has been reported that effect of this component is to dampen out perturbations in other components eventually stabilizing the scheme \cite{pandolfi2001, shen2014}. 
Had it been included, then isolating the effect of parameter $\epsilon^{L,R}$ on physical quantitites would have been difficult.

\noindent For the HLLC scheme, the evolution equation for $\hat{\rho},\hat{p}$
and $\hat{u}$ can be derived and arranged in a matrix form as,

		      \begin{align}
		      \left [ \begin{array}{c}
		      \hat{\rho} \\
		      \hat{u}\\
		      \hat{p}
		      \end{array}\right ]^{n+1} = 
		      \left [ \begin{array}{r}
		      1\\
		      0\\
		      0
		      \end{array}
		      \begin{array}{r}
		      0\\
		      1\\
		      0
		      \end{array}
		      \begin{array}{r}
		     \frac{-2\lambda}{\gamma}\\
		      0\\
		      1-2\lambda
		      \end{array}\right ]
		      \left [ \begin{array}{c}
		      \hat{\rho} \\
		      \hat {u}\\
		      \hat {p}
		      \end{array}\right ]^n
		      \label{eqn:quirkanalysis_hllc}
		      \end{align}
Corresponding evolution equations for the HLLC-SWM scheme can be derived as,  
		      \begin{align}		      
		      \left [ \begin{array}{c}
		      \hat{\rho} \\
		      \hat{u}\\
		      \hat{p}
		      \end{array}\right ]^{n+1} = 
		      \left [ \begin{array}{r}
		      1- \frac{2\lambda\alpha\epsilon}{\sqrt{\gamma}} \\
		      0\\
		      0
		      \end{array}
		      \begin{array}{r}
		      0\\
		      1-\frac{2\lambda\alpha\epsilon}{\sqrt{\gamma}}\\
		      0
		      \end{array}
		      \begin{array}{r}
		      -\frac{2\lambda}{\gamma}\\
		      0\\
		      1-2\lambda(1+\frac{\alpha\epsilon}{\sqrt{\gamma}})
		      \end{array}\right ]
		      \left [ \begin{array}{c}
		      \hat{\rho} \\
		      \hat {u}\\
		      \hat {p}
		      \end{array}\right ]^n
		      \label{eqn:quirkanalysis_hllcswm}		      
		      \end{align}
where $\lambda = \sqrt{\gamma}\frac{\Delta t}{\Delta y}$ denotes a linearized CFL number. The eigenvalues of the coefficient matrices in these cases represent
the amplification factors of the respective perturbation quantities. It is seen from these equations that in case of the HLLC scheme, the amplification factors 
for perturbations in $\hat{\rho},\hat{u},\hat{p}$ are respectively $(1,1,1-2\lambda)$. A von Neumann type stability bound on $\lambda$ can be obtained from these as, 
\begin{align}
  0<\lambda<1 
  \label{eqn:stability_hllc}
\end{align}
It becomes evident that the perturbations in $\rho$ and $u$ are not damped 
by the HLLC scheme while those in $p$ are attenuated for $0<\lambda<1$. These behaviours can be observed in Figs.(\ref{fig:HLLC_density_perturbation},\ref{fig:HLLC_velocity_perturbation},
\ref{fig:HLLC_pressure_perturbation}) that form the right panel of Fig.(\ref{fig:perturbationstudies_hllcswm}). Each plot in these figures
indicate the evolution of perturbation in either density, x-velocity or pressure (indicated by $\hat{\rho}, \hat{u}, \hat{p}$ respectively) for a 
given initial perturbation in one of those quantities (while setting the other initial perturbations to be 0). These plots correspond to a value of $\lambda = 0.2$.
However, as seen from Fig.(\ref{fig:HLLC_pressure_perturbation}), although $\hat{p}$ gets attenuated with time, any finite $\hat{p}$ is repeatedly feeding into 
$\hat{\rho}$ which results in existence of a residual $\hat{\rho}$ that is growing in time. 
These undamped perturbations in $\rho$ and $u$ may eventually amplify leading to unphysical mass flux 
jumps across the shock wave. Such jumps could then force a distortion of the shock profile causing the typical 'bulge' in shock structure and an eventual shock 
instability solution \cite{shen2014}. Since the HLLC scheme is known to produce shock instabilites, it can be inferred from these observations that Quirk's conjecture 
is perfectly valid in its case. 
Gressier et al \cite{gressier2000} have termed schemes with 
atleast one undamped perturbation in physical variable (ie. with non-decaying amplifcation factors) as \textit{marginally stable} schemes. 
By due consideration the HLLC scheme is clearly a \textit{marginally stable} scheme.
 
Examining Eqs.(\ref{eqn:quirkanalysis_hllcswm}) and the left panel of Fig.(\ref{fig:perturbationstudies_hllcswm}) in the light of above arguments, it is 
easy to see why the HLLC-SWM framework will result in a shock stable scheme. First of all we notice that all the three evolution equations of the HLLC-SWM scheme essentially comprise of 
a HLLC component and extra terms that contain the parameters $\epsilon^{L,R}$ and $\alpha$. The amplification factors of the perturbations $\hat{\rho},\hat{u},\hat{p}$ are respectively $\left( 1- \frac{2\lambda\alpha\epsilon^{L,R}}{\sqrt{\gamma}},
1-\frac{2\lambda\alpha\epsilon^{L,R}}{\sqrt{\gamma}},1-2\lambda(1+\frac{\alpha\epsilon^{L,R}}{\sqrt{\gamma}}) \right)$. These lead to the following stability bound on 
$\lambda$, 
\begin{align}
  0<\lambda <\frac{1}{1+\frac{\alpha \epsilon^{L,R}}{\sqrt{\gamma}}}
  \label{eqn:stability_hllemswm}
\end{align}

For a $\lambda$ in these bounds, it can be seen that any value for $\epsilon^{L,R},\alpha >0$ will introduce a damping factor of 
$\frac{2\lambda\alpha\epsilon^{L,R}}{\sqrt{\gamma}}$ into each of these perturbations. Notice that while this additional factor damps $\hat{u}$, it provides 
an ancillary damping to $\hat{p}$ over the preexisting damping coefficient $(1-2\lambda)$. In case of $\hat{\rho}$, the additional damping factor 
not only ensures that independent perturbations in density are attenuated, but also that any residual perturbations 
arising in $\hat{\rho}$ due to action of $\hat{p}$ are also suppressed. Thus in general, the additional factor ensures that the perturbations in all the three 
primitive quantities die down in time. 
These behaviours can be clearly observed in left panel of Fig.(\ref{fig:perturbationstudies_hllcswm}). The plots correspond to $\lambda = 0.2$ and 
$\alpha,\epsilon^{L,R} =1.0$. Given these damping characterisitcs, 
the HLLC-SWM scheme is very unlikely to produce variation of mass flux in the vicinity of the shock profile. This would indicate that numerical solutions of 
the HLLC-SWM scheme would remain shock instability free \cite{shen2014}. Such a scheme with decaying amplification factors are 
termed as \textit{strictly stable }scheme and are known to be shock instability free \cite{gressier2000}. Thus modified dissipation $\overline{\mathbf{D}}_{HLL}$
proposed in Eq.(\ref{eqn:HLLCSWMdissipation-absoluteform}) thus helps convert the HLLC scheme from 
its inherent \textit{marginally stable} behaviour to a \textit{strictly stable} behaviour.

	    \begin{figure}[H]
		\begin{subfigure}[b]{0.5\textwidth}
		  \begin{center}
		 \includegraphics[width=5cm,height=4.5cm]{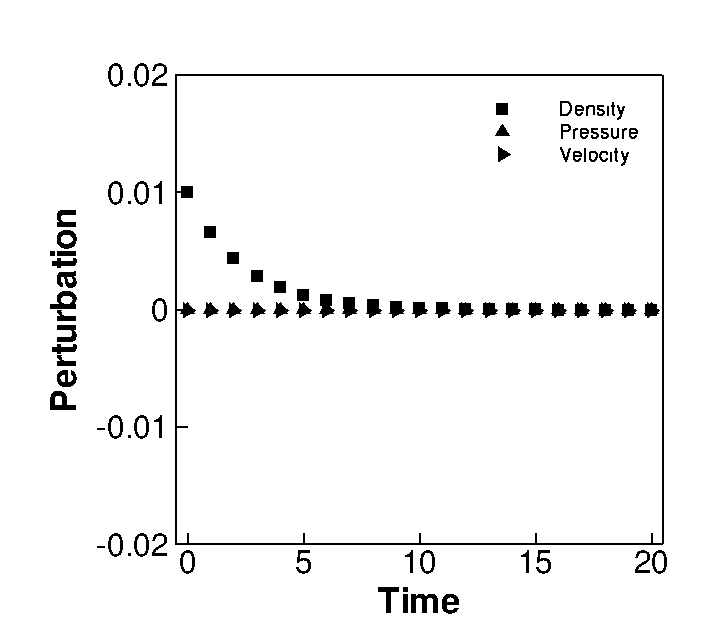}
		  \caption{HLLC-SWM $\left(\hat{\rho} = 0.01, \hat{u} = 0, \hat{p} = 0\right)$  }
		  \label{fig:HLLCWM_density_perturbation}
		  \end{center}
		\end{subfigure}
		\begin{subfigure}[b]{0.5\textwidth}
		  \includegraphics[width=5cm,height=4.5cm]{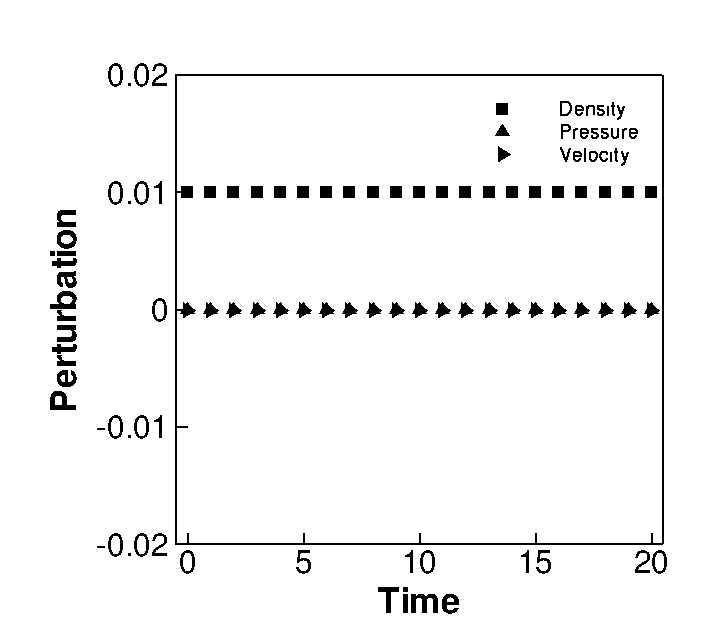}
		  \caption{HLLC $\left(\hat{\rho} = 0.01, \hat{u} = 0, \hat{p} = 0\right)$}
		  \label{fig:HLLC_density_perturbation}
		  \end{subfigure}
	    \end{figure}
		
	    \begin{figure}[H]
		\ContinuedFloat
		 \begin{subfigure}[b]{0.5\textwidth}
		  \includegraphics[width=5cm,height=4.5cm]{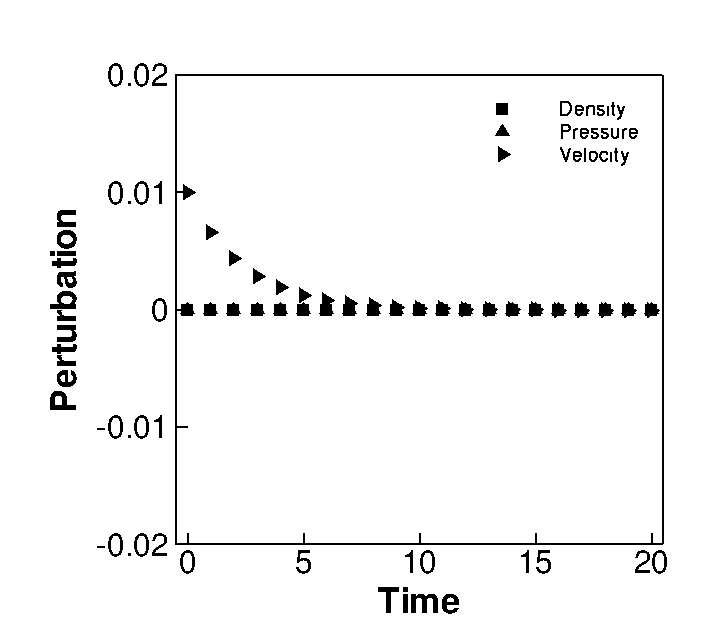}
		  \caption{HLLC-SWM $\left( \hat{\rho} = 0, \hat{u} = 0.01, \hat{p} = 0 \right)$}
		  \label{fig:HLLCWM_velocity_perturbation}
		 \end{subfigure}
		 \begin{subfigure}[b]{0.5\textwidth}
		  \includegraphics[width=5cm,height=4.5cm]{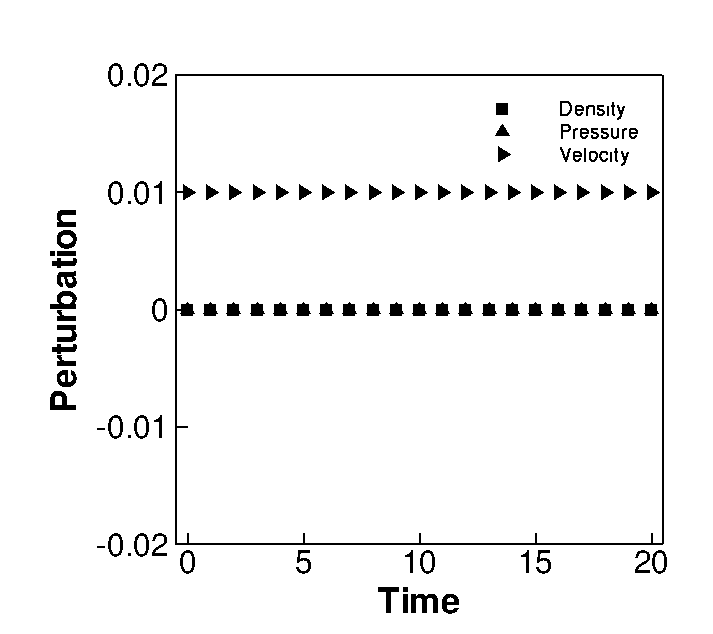}
		  \caption{HLLC $\left(\hat{\rho} = 0, \hat{u} = 0.01, \hat{p} = 0\right)$}
		  \label{fig:HLLC_velocity_perturbation}
		 \end{subfigure}
		 \begin{subfigure}[b]{0.5\textwidth}
		  \includegraphics[width=5cm,height=4.5cm]{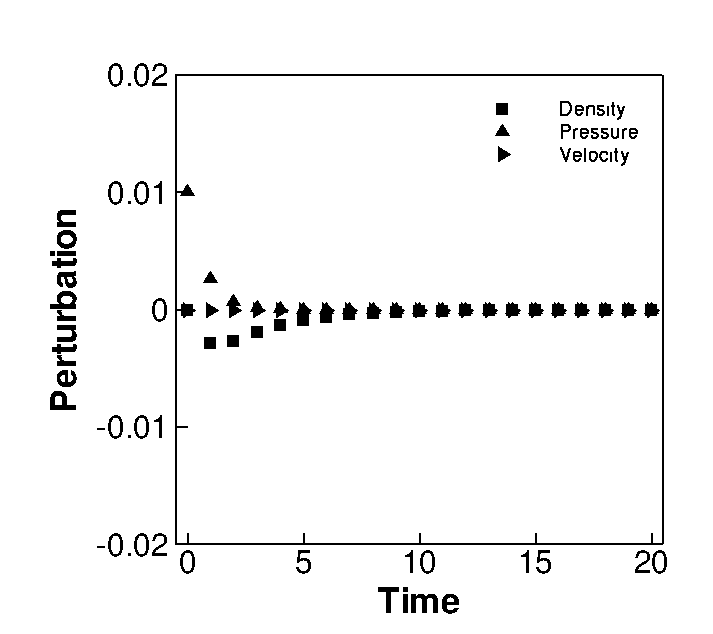}
		  \caption{HLLC-SWM $\left(\hat{\rho} = 0, \hat{u} = 0, \hat{p} = 0.01\right)$}
		  \label{fig:HLLCWM_pressure_perturbation}
		 \end{subfigure}
		 \begin{subfigure}[b]{0.5\textwidth}
		  \includegraphics[width=5cm,height=4.5cm]{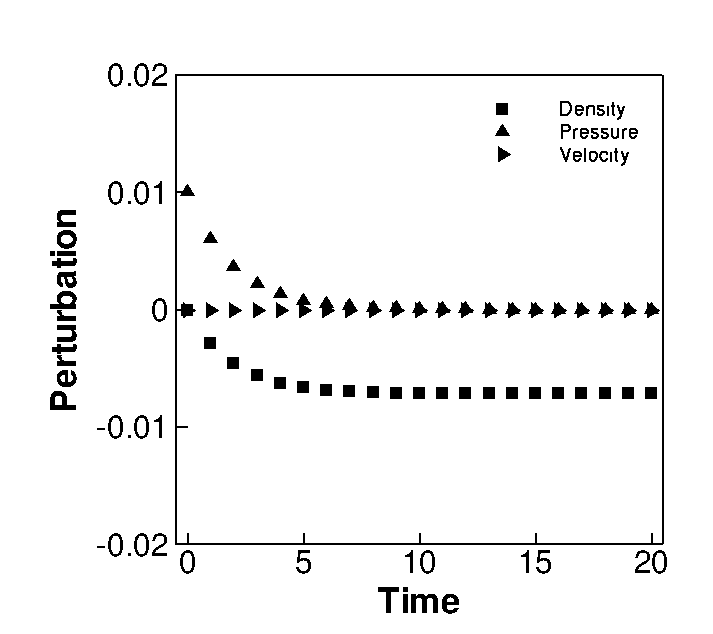}
		  \caption{HLLC $\left(\hat{\rho} = 0, \hat{u} = 0, \hat{p} = 0.01\right)$}
		  \label{fig:HLLC_pressure_perturbation}
		 \end{subfigure}
	    \caption{Comparison of evolution of density, x-velocity and pressure perturbations in the HLLC-SWM and the HLLC schemes.}
	    \label{fig:perturbationstudies_hllcswm}
	    \end{figure}

\noindent Although various authors \cite{sanders1998,peery1988,lin1995} have successfully used eigenvalue modification for the Roe scheme to cure shock 
instabilities it is not entirely made clear as to how these modifications affected the evolution of errors in physical variables during a computation to eventually secure a
shock stable solution. 
The analysis presented here clarifies this in the context of the HLL based schemes. As seen from Eq.(\ref{eqn:quirkanalysis_hllcswm}), the wavespeed modification suggested 
in Eq.(\ref{eqn:hllc_modifiedwavespeedestimates}) for the HLLC scheme
expressed in the form shown in Eq.(\ref{eqn:hllc-modifiedform})
directly affects the amplification factors that control the growth or decay of perturbations in physical quantities. A damping of these perturbations is assured through
the dissipation parameters $\epsilon^{L,R}$ and $\alpha$ which eventually promises shock stability. 

In the next section, we discuss the some strategies to estimate the quantity of the dissipation parameter $\epsilon^{L,R}$ to be used while calculating the HLLC-SWM flux at any interface.

\section{Estimation of dissipation parameter $\epsilon^{L,R}$}
\label{sec:estimationofepsilon}
Several choices for estimation of $\epsilon^{L,R}$ can be designed. Since the purpose of $\epsilon^{L,R}$ is primarily to increment the value of $\overline{S_L}$ and 
$\overline{S_R}$ locally in the vicinity of a shock wave and tend to zero elsewhere, any viable shock sensor is a good candidate for it.
Since shock instability is primarily a multidimensional phenomenon, it is also advisable to have an estimation of $\epsilon^{L,R}$ that takes this fact into account 
\cite{sanders1998}. Below we discuss some strategies to achieve a satisfactory estimate of $\epsilon^{L,R}$ based 
on these considerations. 

\subsection{Eigenvalues based estimation of $\epsilon^{L,R}$}
\label{sec:characteristicbasedestimationofepsilon}

\begin{figure}[ht]
	    \centering
	    \includegraphics[scale=0.25]{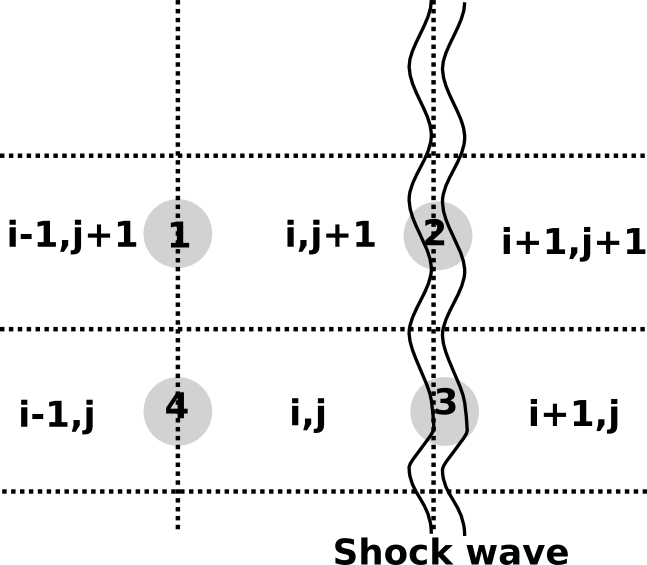}
	   \caption{Stencil adopted for calculating $\eta$'s required for estimation of $\epsilon^{L,R}$ at ($i,j+1/2$) interface.}
	   \label{fig:stencilforepsilon}
	    \end{figure}

One possible estimate for $\epsilon^{L,R}$ can be based upon the idea of an entropy fix. Several authors \cite{pandolfi2001,sanders1998,peery1988,lin1995} have used such 
entropy fix based estimates to save the Roe scheme from shock instabilities. For a typical cell interface $(i,j+1/2)$, $\epsilon^{L,R}$ may be defined as,
\begin{align}
 \epsilon_{i,j+1/2}^{L,R} = max(\eta_1,\eta_2,\eta_3,\eta_4)
\end{align}
where each $\eta_k$ ($k$=1...4) corresponds to the dissipation quantity calculated at predefined interfaces that form a stencil around $(i,j+1/2)$ as shown 
in Fig.(\ref{fig:stencilforepsilon}). 

Each $\eta_k$ can be 
evaluated using a common entropy fix formula,
\begin{align}
 \eta_k = \frac{1}{2}max( |\lambda_p(\mathbf{U}_R) - \lambda_p(\mathbf{U}_L)| )
\end{align}
where $p$ iterates over all the characteristic wavespeeds $\lambda_p$ of the Riemann problem at $k^{th}$ interface. An $\epsilon^{L,R}$ at any interface defined 
in the above form represents the largest one dimensional entropy correction of all the associated interfaces considered around it.
As pointed out in \cite{pandolfi2001} this definition is designed carefully to avoid extraneous 
dissipation from being evaluated when computing flow features other than shocks with naturally occuring gradients; like the boundary layers. This is primarily
achieved by excluding the $({i,j+1/2})^{th}$ interface itself from the stencil used to calculate $\epsilon^{L,R}$. 
Further, it is important to state here that the idea of an entropy fix is utilized in the above strategy only to determine 
an estimate for $\epsilon^{L,R}$ that is $\mathcal{O} (S_L,S_R)$. Since the HLLC-SWM framework builds upon the HLLC scheme which by default is entropy satisfying, 
it does not require fixes to ensure entropy satisfying solutions. 
It is well known from the works of many authors \cite{xu2001,sanders1998,ren2003,shen2014,nishikawa2008,liou2000} that one of the 
reasons for shock instability appearing in solutions of contact and shear preserving schemes has to do with its inability to provide adequate multidimensional dissipation 
in the vicinity of a shock front. The above definition of $\epsilon^{L,R}$ along with modification proposed in Eq.(\ref{eqn:hllc_modifiedwavespeedestimates}) 
may be considered as an effective strategy to address this concern in case of the HLLC scheme.
The dissipation provided through $\epsilon^{L,R}$ has a multidimensional flavor because the dissipation introduced at a longitudinal interface is a function of the solution variation in the 
transverse direction and vice versa \cite{sanders1998}. An HLLC-SWM framework that uses this estimate for $\epsilon^{L,R}$ is hereby termed HLLC-SWM-E (\textbf{E}igenvalue based) 
scheme.

\subsection{Pressure based estimation of $\epsilon^{L,R}$}
\label{sec:pressurebasedestimationofepsilon}

Although quite appealing, the entropy fix based estimation of $\epsilon^{L,R}$ discussed above may cause certain loss of accuray in shear dominated viscous flows.
This is due to the fact that $\epsilon^{L,R}$ in this definition is employed both as a shock sensor and as an estimate of the quantity of dissipation to be provided at it.
Thus, in flows with strong velocity gradients, the value of $\epsilon^{L,R}$, instead of tending 
towards zero within the shear layers, could become a non zero quantity thereby introducing erroneous dissipation into the scheme and compromising the accuracy of 
the solution. An interesting alternative to this would be to use seperate functions for shock detection and estimation of dissipation quantity. This can be done by
including a pressure based shock sensor in the definition of $\epsilon^{L,R}$. The advantage of pressure
based sensor is that they become active at shocks and remain inactive in viscous dominated flows and inviscid contacts.
Hence we define a new
$\epsilon^{L,R}$ through a pressure based shock sensor as,

\begin{align}
 \epsilon_{i,j+1/2}^{L,R} = (1-\omega_{i,j+1/2})\ \ max(\eta_1,\eta_2,\eta_3,\eta_4)
 \label{eqn:epsilonbasedonpressure}
\end{align}
where $\omega_{i,j+1/2}$ is a pressure based shock sensor that can be constructed as \cite{zhang2017},
\begin{align}
 \omega_{i,j+1/2} = min_k (f_k),\ \ \ k=1...4
 \label{eqn:definitionofomega}
\end{align}
where $f_k$'s are pressure ratio based functions evaluated on a predefined stencil around $({i,j+1/2})$ interface shown in Fig.(\ref{fig:stencilforepsilon}). 
At $k^{th}$ interface, $f_k$ is defined as,  
\begin{align}
 f_k = min\left ( \frac{p_R}{p_L},\frac{p_L}{p_R} \right )_k^\beta
 \label{eqn:definitionofpressureratiofucntion}
\end{align}
Here $p_R$ and $p_L$ denotes the right and left cell center pressures across $k^{th}$ interface. A value of $\beta=5.0$ is used. 
Note that this pressure based $\epsilon^{L,R}$ also retains the merits of the shock sensor discussed in 
Sec.(\ref{sec:characteristicbasedestimationofepsilon}) ie, a multidimensional stencil
and local solution dependence. An HLLEM-SWM scheme that uses this kind of $\epsilon^{L,R}$ will hereby be addressed as HLLC-SWM-P (\textbf{P}ressure based) scheme.
In the next section, we complete the description of HLLC-SWM framework by estimating an appropriate value for the parameter $\alpha$. 

\section{Estimation of dissipation parameter $\alpha$}
\label{sec:estimationofalpha}

The dissipation parameter $\alpha$ controls the extent to which the wave speeds $S_L$ and $S_R$ are modified by the $\epsilon^{L,R}$ parameter. Too large a value of $\alpha$ could
result in larger magnitudes of $\overline{S_L}$ and $\overline{S_R}$ which as seen from Eq.(\ref{eqn:dissipationcoefficients_simplified}) will result in larger magnitude of
$\overline{\mathbf{D}}_{HLL}$ and eventually threaten to spoil the accuracy of the solution. On the other hand, too small a value could fail to provide adequate numerical disspation 
to counter the threat of shock instability. Thus, choice of $\alpha$ plays a crucial role in the overall efficiency of the HLLC-SWM class of schemes and hence must be carefully chosen.
We envision that a reasonable lower bound on $\alpha$ that is capable of dealing with shock instabilities can be obtained by studying the effect of varying $\alpha$ on the stability 
of a shock profile of a simple shock instability test problem: the steady simulation of a two dimensional thin shock. Ideally, the steady state solution of this 
problem will consist of the original unperturbed shock profile that will be preserved forever. However, presence of shock instability
would immediately manifest as unsteady perturbations in physical quantities in the aftershock region.
Instead of directly resorting to cumbersome numerical simulations to achieve this objective, we utilize the technique of matrix based stability analysis as proposed by 
Dumsber et al \cite{dumbser2004}. The present work chooses a $M=7$ two dimensional thin steady shock to be located between the 6th and 7th cells of a regular Cartesian grid comprising of 11 $\times$ 11 
cells on a 1.0$\times$
1.0 domain. The initial conditions on the supersonic side of the shock are chosen as $(\rho,u,v,p)_L=(1.0,1.0,0.0,0.01457)$ while Rankine-Hugoniot 
conditions are used to determine the corresponding conditions on the subsonic side. Periodic conditions are applied on the top and bottom boundaries. Since the HLLC scheme exactly preserves a thin shock, 
the initial and boundary conditions can be exactly specified \cite{dumbser2004}. The eigenvalue problem is solved using the eigenvalue
function \textit{eig()} from \textit{linalg} package in python version 2.7.6.

\subsection{Estimation of dissipation parameter $\alpha$ for the HLLC-SWM-E scheme}
\label{sec:estimationofalphaforhllcswm_E}
\begin{figure}[H]
		\begin{subfigure}[b]{0.5\textwidth}
		  \begin{center}
		 \includegraphics[width=5.5cm,height=5.5cm]{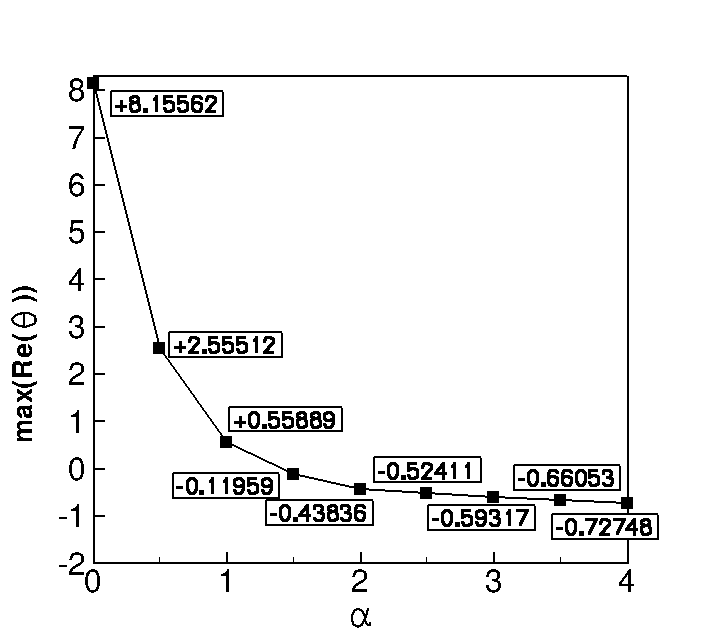}
		  \caption{}
		  \label{fig:effectofalpha_hllc_swm_e}
		  \end{center}
		\end{subfigure}
		\begin{subfigure}[b]{0.5\textwidth}
		  \begin{center}
		  \includegraphics[width=5.5cm,height=5.5cm]{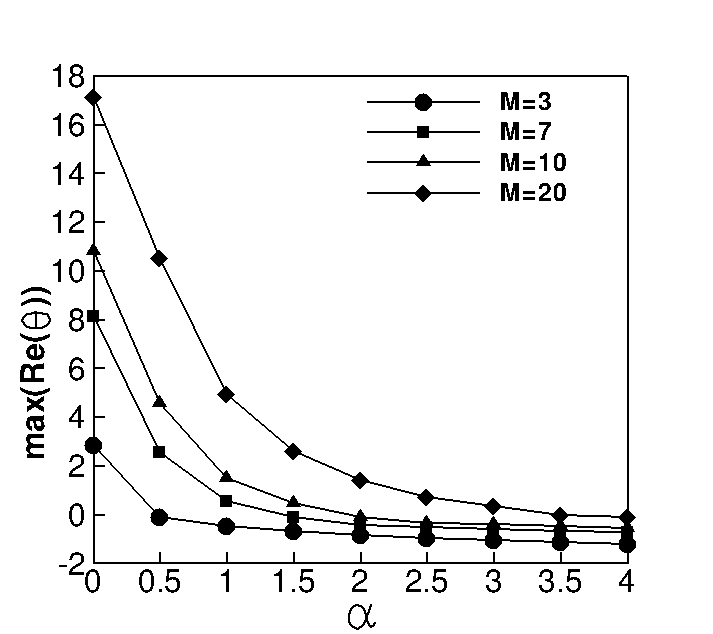}
		  \caption{}
		  \label{fig:effectofalphaforfourMachnumbers_hllc_swm_e}
		  \end{center}
		  \end{subfigure}
	    \caption{(a)Plot showing the effect of $\alpha$ on $max(Re(\theta))$ for a $M=7$ stationary shock computed using the HLLC-SWM-E scheme (b) Plot showing the effect of $\alpha$ on $max(Re(\theta))$ for varying inflow Mach numbers M$=3,7,10$ and $20$ of two dimensional stationary shock problem 
	   computed using the HLLC-SWM-E scheme.}
	    \label{fig:matrixanalysisresults}
	    \end{figure}

In this section we estimate the value of $\alpha$ that ensures a shock stable HLLC-SWM-E scheme. 
Fig.(\ref{fig:effectofalpha_hllc_swm_e})
shows the variation of the max Real part of eigenvalues $\theta$ (indicated in the boxes) as a function of varying $\alpha$
values computed using the HLLC-SWM-E scheme.


	    
\noindent Since $max(Re(\theta))$ represents the largest error growth rate, all $max(Re(\theta))$ value in Fig.(\ref{fig:effectofalpha_hllc_swm_e}) 
lying above 0.0 corresponds to unstable (exponentially amplifying) solutions  while those lying below represents shock stable 
(exponentially decaying) solutions on
this problem. An $\alpha=0.0$ corresponds to the unmodified HLLC scheme which is known to be unstable on this problem. 
The positive $max(Re(\theta))$ value of +8.15562 
associated with $\alpha=0.0$ confirms that the numerical errors have a propensity to grow and eventualy spoil the solution. It is interesting to note that even $\alpha=1$ does not ensure complete 
shock stability for the HLLC-SWM-E scheme. This means that unlike suggestions in \cite{pandolfi2001,sanders1998}, for the Roe scheme, where an estimate of $\epsilon^{L,R}$ discussed
in Sec.(\ref{sec:characteristicbasedestimationofepsilon}) was claimed to ensure shock instability, an $\alpha>1.0$ is necessary to save the HLLC scheme.
Further, it is interesting to note that beyond $\alpha=1.0$, a slight increase in $\alpha$ is seen to stabilize the solution in the linear limit. 
With increasing $\alpha$, the maximum error growth rate is found to drop indicating a tendency to establish a shock stable solution.  

\begin{figure}[H]
		\begin{subfigure}[b]{0.5\textwidth}
		  \begin{center}
		 \includegraphics[width=4cm,height=4cm]{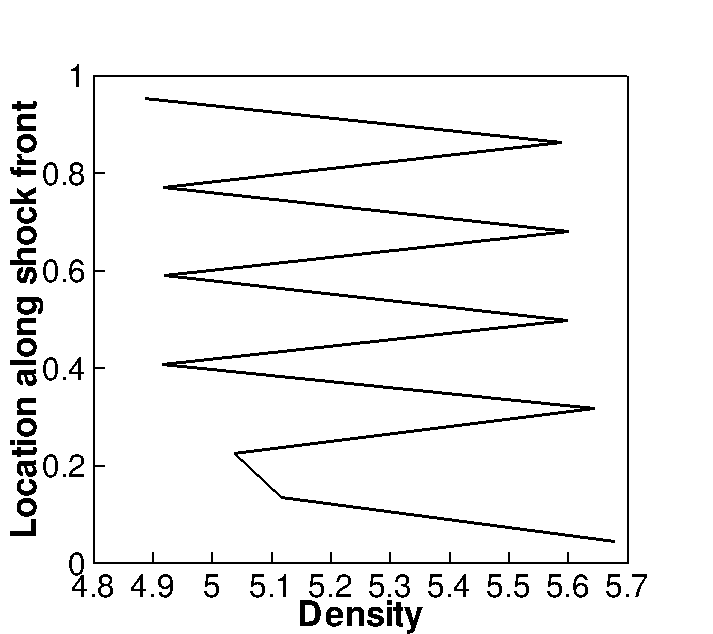}
		  \caption{$\alpha=0.0$}
		  \label{fig:hllc_aftershockdensity}
		  \end{center}
		\end{subfigure}
		 \begin{subfigure}[b]{0.5\textwidth}
		  \begin{center}
		  \includegraphics[width=4cm,height=4cm]{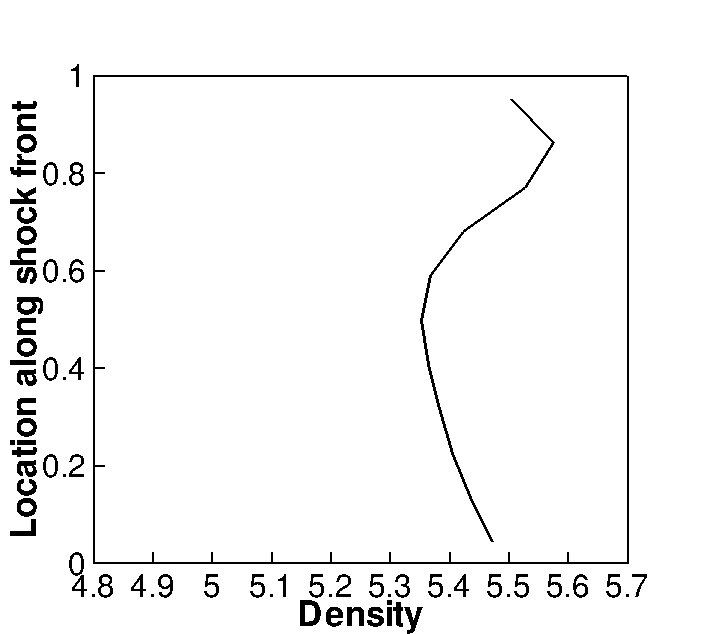}
		  \caption{$\alpha=1.0$}
		  \label{fig:hllcswm_alpha_1_aftershockdensity}
		  \end{center}
		 \end{subfigure}
		 \begin{subfigure}[b]{0.5\textwidth}
		  \begin{center}
		  \includegraphics[width=4cm,height=4cm]{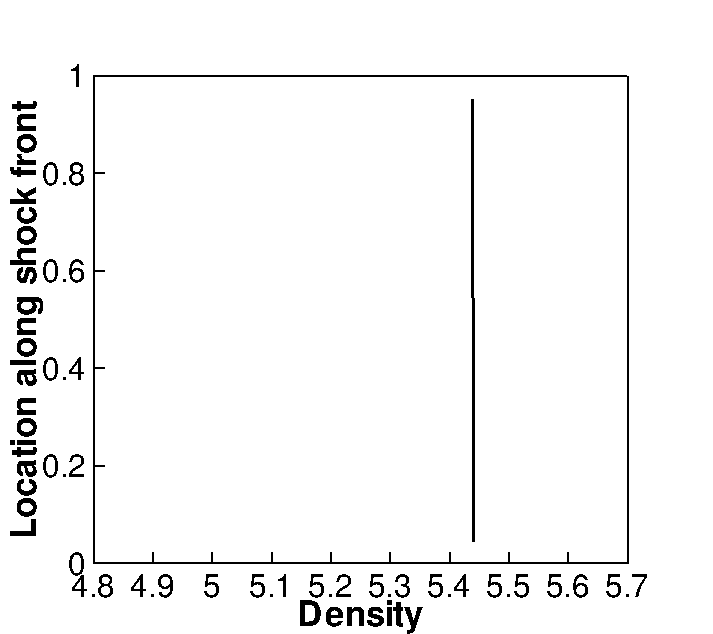}
		  \caption{$\alpha=2.0$}
		  \label{fig:hllcswm_alpha_2_aftershockdensity}
		  \end{center}
		 \end{subfigure}
		\begin{subfigure}[b]{0.5\textwidth}
		 \begin{center}
		  \includegraphics[width=4cm,height=4cm]{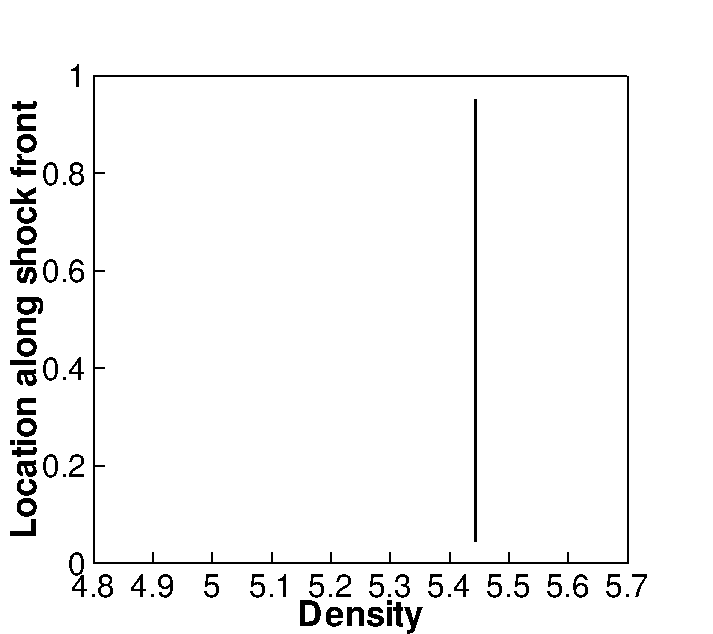}
		  \caption{$\alpha=3.5$}
		  \label{fig:hllcswm_alpha_3pt5_aftershockdensity}
		  \end{center}
		 \end{subfigure}
		 \caption{Variation of density along the computed shock front extracted from the 7th cell. The solution was obtained using the proposed HLLC-SWM-E scheme 
	    for varying $\alpha$ values. The deviation of value of density variable from its exact steady value of 5.44 indicates presence of instabilities. The solutions are obtained at
	    t=20 units.}
	    \label{fig:densityaftershock_hllcswm_e}
	   \end{figure}

We verify these predictions in case of the HLLC-SWM-E scheme using a numerical 
simulation of the above problem carried out using a Finite volume first order accurate Euler code.
Fig.(\ref{fig:densityaftershock_hllcswm_e}) shows plots of cell centered density values versus the y locations extracted right behind the 
shock, ie. from the 7th coloum of cells in the 
computational domain. Since the problem has an exact steady state solution, any deviation of value of the density variable from its exact steady value of 5.44 clearly 
indicates presence of instabilities. The solutions are obtained at t=20 units.

\noindent Fig.(\ref{fig:hllc_aftershockdensity}) clearly shows the saw tooth profiled density variation that is typical of a shock unstable solution \cite{quirk1994}. 
These undulations in physical variables may enhance or 
decay depending on parameters like grid aspect ratio or inflow Mach number of the specific problem considered. Further, as predicted by the analysis, increasing $\alpha$ values 
are seen to stabilize the solution. For $\alpha \geq2.5$ the HLLC-SWM-E scheme is able to infuse sufficient dissipation to render a shock stable solution on this
problem. However as reported in \cite{dumbser2004}, inflow Mach number can have a predominant effect on the proliferation of these instabilities.
Hence, a choice of $\alpha$ 
that is capable of dealing with instabilities over a wide range of Mach numbers is preferred. To include the effect of Mach numbers, the above analysis is repeated 
for three additional Mach numbers; M=3, M=10 and M=20. Fig.(\ref{fig:effectofalphaforfourMachnumbers_hllc_swm_e}) shows the 
variation of $max(Re(\theta))$ with respect to $\alpha$ for these additional Mach numbers.

\noindent It is seen from Fig.(\ref{fig:effectofalphaforfourMachnumbers_hllc_swm_e}) that in general, higher Mach numbers corresponds to larger error growth rates for any value of $
\alpha$. For $\alpha=0.0$ which corresponds to the HLLC scheme, increase in Mach numbers leads to a drastic decrease in theoretical shock stability. This is an expected
behaviour from a typical shock instability prone scheme like the HLLC \cite{dumbser2004}. Further at any given Mach number, an increasing value of $\alpha$ is seen to 
reduce the error growth rate. In case of the HLLC-SWM-E scheme, 
each Mach number has a corresponding $\alpha$ value at which the $max(Re(\theta))$ crosses over into the stable region. From Fig.(\ref{fig:effectofalphaforfourMachnumbers_hllc_swm_e})
it is seen that choice of $\alpha=3.5$ favors a linearly stable solution over the range of Mach numbers investigated.

\subsection{Estimation of dissipation parameter $\alpha$ for the HLLC-SWM-P scheme}
\label{sec:estimationofalphaforhllcswm_P}
\begin{figure}[h]
		  \begin{center}
		 \includegraphics[width=5.5cm,height=5.5cm]{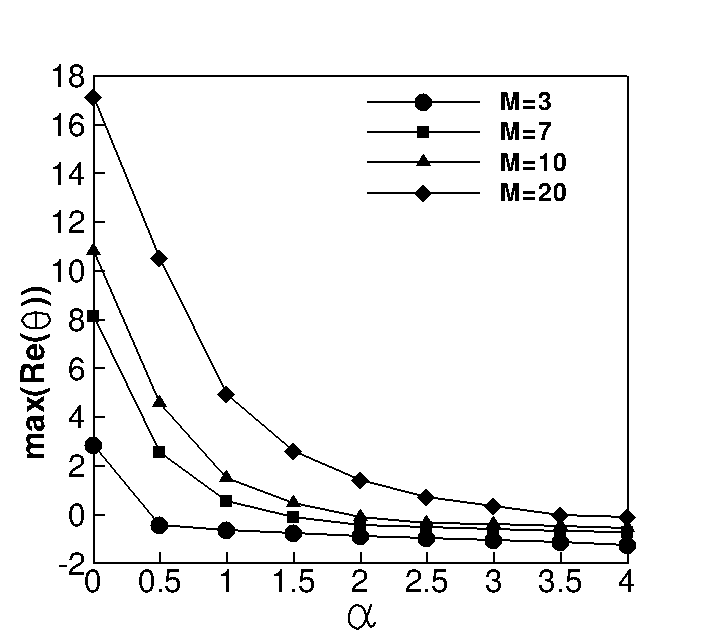}
		  \caption{Plot showing the effect of $\alpha$ on $max(Re(\theta))$ for varying inflow Mach numbers M$=3,7,10$ and $20$ of two dimensional stationary shock problem 
	   computed using the HLLC-SWM-P scheme.}
		  \label{fig:effectofalphaforfourMachnumbers_hllc_swm_p}
		  \end{center}
		\end{figure}

Fig.(\ref{fig:effectofalphaforfourMachnumbers_hllc_swm_p}) shows the variation of $max(Re(\theta))$ with $\alpha$ for the stationary shock problem computed using the HLLC-SWM-P
scheme. It is evident that the behaviour of the stability region for the HLLC-SWM-P scheme is quite similar to that of the HLLC-SWM-E scheme. A value of $\alpha=3.5$ is seen to
assure theoretical shock stability. We have confirmed this prediction with numerical experiments similar to that done in Sec.(\ref{sec:estimationofalphaforhllcswm_E}). 

\noindent We present the eigenvalue spectrums for M=7 stationary normal shock computed using the HLLC-SWM schemes in Fig.(\ref{fig:matrixanalysisresults}).
An $\alpha=3.5$ is used for both schemes. Corresponding specturm for the original HLLC is also provided for comparison. 
The eigenvalue spectrum visibly shows how the proposed HLLC-SWM-E and the HLLC-SWM-P schemes utilize the modified dissipation $\overline{\mathbf{D}}_{HLL}$ 
to shift the unstable error growth rates to the stable region. While the HLLC scheme has the largest error growth rate of +8.15562, the HLLC-SWM-E and the HLLC-SWM-P schemes
with  $\alpha=3.5$ are able to bring this 
down to -0.66053 and -0.59317 respectively. Amongst the HLLC-SWM schemes, the HLLC-SWM-E scheme turns out to be more stable. Further, it is important to note that the shock stability predicted by the matrix analysis for the proposed HLLC-SWM schemes 
are independant of the time discretization technique adopted or the CFL number employed because only spatial discretization of the governing equations has been used to construct the stability matrix. 

	     \begin{figure}[!]
		\begin{subfigure}[b]{0.5\textwidth}
		  \begin{center}
		 \includegraphics[width=5cm,height=5cm]{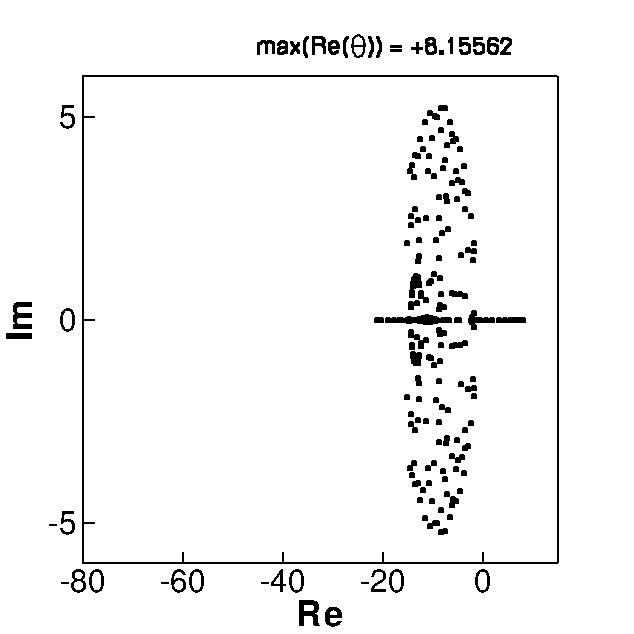}
		  \caption{HLLC}
		  \label{fig:hllc_matrixanalysis}
		  \end{center}
		\end{subfigure}
		\begin{subfigure}[b]{0.5\textwidth}
		  \begin{center}
		  \includegraphics[width=5cm,height=5cm]{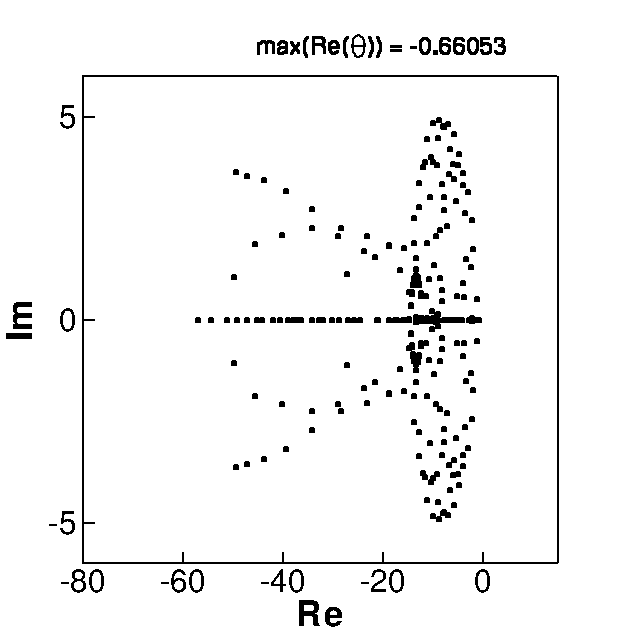}
		  \caption{HLLC-SWM-E}
		  \label{fig:hllcswm_e_matrixanalysis}
		  \end{center}
		  \end{subfigure}
		  \begin{subfigure}[b]{0.5\textwidth}
		  \begin{center}
		  \includegraphics[width=5cm,height=5cm]{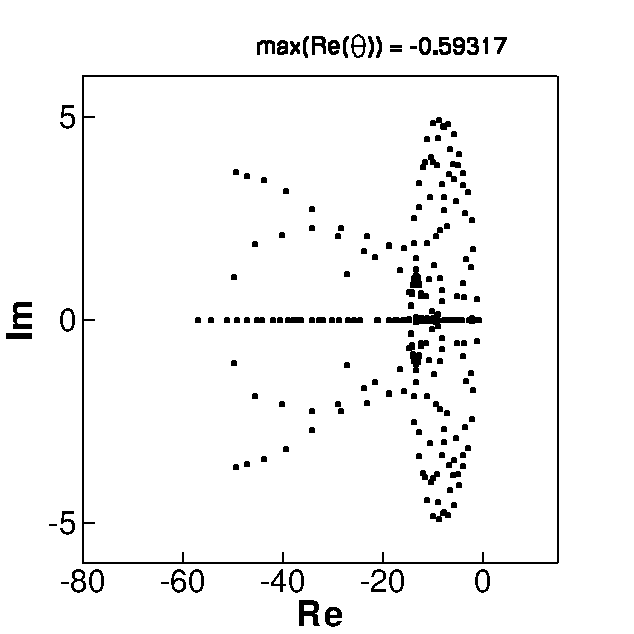}
		  \caption{HLLC-SWM-P}
		  \label{fig:hllcswm_p_matrixanalysis}
		  \end{center}
		  \end{subfigure}
	    \caption{ Eigenvalue spectrums for  M=7 stationary shock problem obtained from matrix analysis (a) HLLC scheme (b) HLLC-SWM-E scheme (c) HLLC-SWM-P scheme. 
	   $\alpha$ taken to be 3.5 for both the HLLC-SWM-E and the HLLC-SWM-P schemes.}
	    \label{fig:matrixanalysisresults}
	    \end{figure}

\section{Numerical verification of the HLLC-SWM schemes}
\label{sec:numericaltests}
In this section, we demonstrate the efficacy of the proposed HLLC-SWM schemes through a series of strict test problems on which the original HLLC scheme
is known to produce shock unstable results. It has been reported in \cite{gressier2000} that
the intensity of instability manifestation in certain test problems reduces as the order of accuracy of the computation increases; although it does not completely remove it.
Based on this, certain test cases are computed as plain first order while second order solutions are sought for others to clearly showcase the improvement in the 
computed flow field using the proposed schemes. Second order spatial accuracy in 
primitive variables is achieved by limiting gradients obtained using
Green Gauss method \cite{balzek2005} with Barth Jersperson limiter \cite{barth1989}. Second order time accuracy is achieved using 
strong stability preserving variant of the Runge Kutta method \cite{Gottlieb2001}. All boundary conditions are set using ghost cells.
A value of $\alpha =3.5$ is used to configure both the HLLC-SWM schemes.

\subsection{Positivity tests}
Batten et al \cite{batten1997} has proved analytically that the original HLLC scheme is a positively conservative
scheme and that it passes the tough numerical test cases found in \cite{einfeldt1991}. 
This section investigates whether the 
modification of nonlinear wavespeeds proposed in Eq.(\ref{eqn:hllc_modifiedwavespeedestimates}) adversely affects this important property of the HLLC scheme. 
Computations are done on 100$\times$50 
structured Cartesian mesh on a domain of size 1.0$\times$1.0. First order solutions are sought and the CFL is chosen as 0.5. 
The first test case is a Riemann problem with initial conditions $(\rho,u,v,p)_L=(1.0,-2.0,0.0,0.4)$ and $(\rho,u,v,p)_R=(1.0,2.0,0.0,0.4)$. The 
boundaries are maintained as open using simple extrapolation. The 
solution to this problem describes two receding expansion waves that originate from the center of the domain and move toward the boundary. 
%
	    \begin{figure}[H]
		\begin{subfigure}[b]{0.5\textwidth}
		  \begin{center}
		 \includegraphics[width=5.5cm,height=5cm]{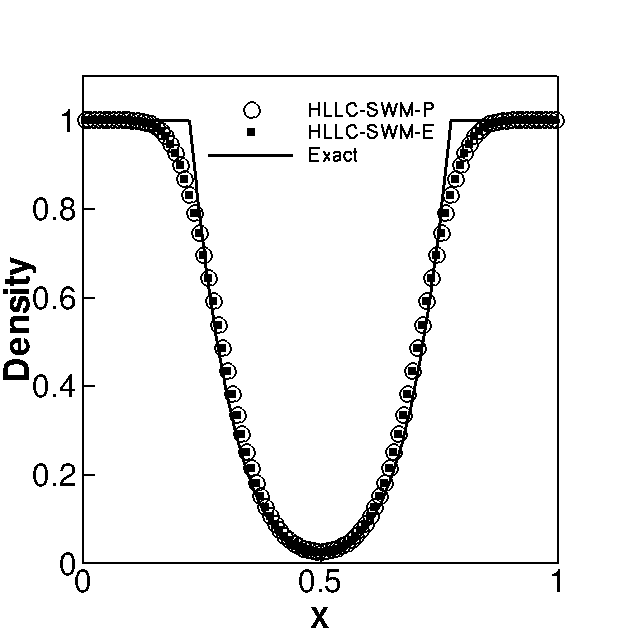}
		  \caption{Density}
		  \label{fig:positivity_test_1_density}
		  \end{center}
		\end{subfigure}
		\begin{subfigure}[b]{0.5\textwidth}
		  \begin{center}
		  \includegraphics[width=5.5cm,height=5cm]{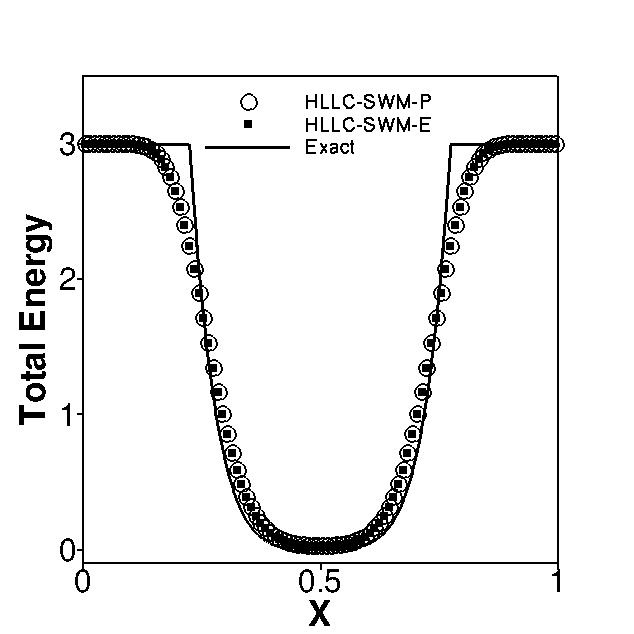}
		  \caption{Total energy}
		  \label{fig:positivity_test_1_energy}
		  \end{center}
		  \end{subfigure}
	   \caption{Solution to the receding expansion wave problem computed by the HLLC-SWM schemes shown at $t=0.1$.}
	    \label{fig:positivity_test_1}
	    \end{figure}  
\noindent Fig.(\ref{fig:positivity_test_1}) shows the result of this test case where both density and total
energy are plotted along the length of the tube after $t=0.1$. Exact solution to this problem can be computed using the
exact Riemann solver of Godunov \cite{godunov1959} and is also plotted in the same figure for reference. 
It is clearly seen that both HLLC-SWM schemes are able to compute the solution without failure and admits physically acceptable density and 
total energy profiles.

\begin{figure}[H]
		\begin{subfigure}[b]{0.5\textwidth}
		  \begin{center}
		 \includegraphics[width=5.5cm,height=5cm]{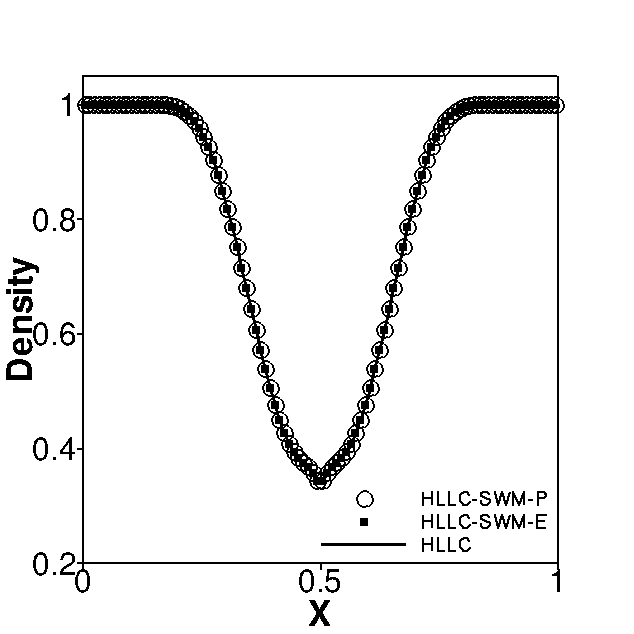}
		  \caption{Density}
		  \label{fig:positivity_test_2_density}
		  \end{center}
		\end{subfigure}
		\begin{subfigure}[b]{0.5\textwidth}
		  \begin{center}
		  \includegraphics[width=5.5cm,height=5cm]{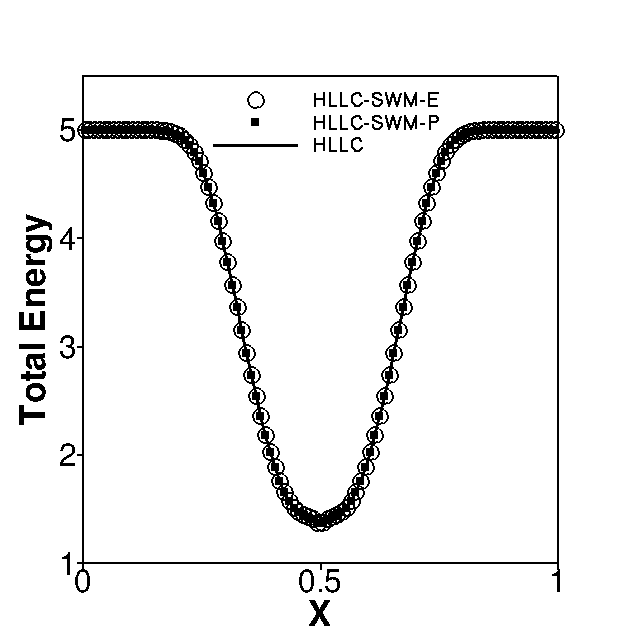}
		  \caption{Total energy}
		  \label{fig:positivity_test_2_energy}
		  \end{center}
		  \end{subfigure}
	    \caption{Solution to the second Riemann problem that consists of two receding expansion waves in x direction seperated by a shear flow in y-direction. The solution shown 
	    for the schemes are obtained at $t=0.1$.}
	    \label{fig:positivity_test_2}
	    \end{figure}

The second Riemann problem is initialized using the condition $(\rho,u,v,p)_L=(1.0,-1.0,-2.0,5.0)$ and $(\rho,u,v,p)_R=(1.0,1.0,2.0,5.0)$. The solution 
of this problem consists of a pair of receding expansion fans seperated by a shearing flow in y-direction.
Fig.(\ref{fig:positivity_test_2}) shows the solution to this problem computed using the HLLC-SWM schemes shown at $t=0.11$. For reference, the 
solution computed by the HLLC scheme at the same time is also provided in Fig.(\ref{fig:positivity_test_2}). It is seen that the HLLC-SWM 
schemes are able to compute the solution to this problem too without failing. The solutions computed by the HLLC-SWM schemes are atleast as accurate as the HLLC scheme.
Solutions to these demanding test problems suggest that the modification proposed in 
Eq.(\ref{eqn:hllc_modifiedwavespeedestimates}) seems to have no adverse effect 
on the positivity property of the HLLC solver and that both HLLC-SWM schemes are positively conservative in nature. It is also observed that 
choice of value of $\alpha$ does not affect this property. 
%

\subsection{Moving shock instability problem (Inviscid)}
\begin{figure}[H]
		\begin{subfigure}[b]{0.5\textwidth}
		  \begin{center}
		 \includegraphics[width=6cm,height=2.5cm]{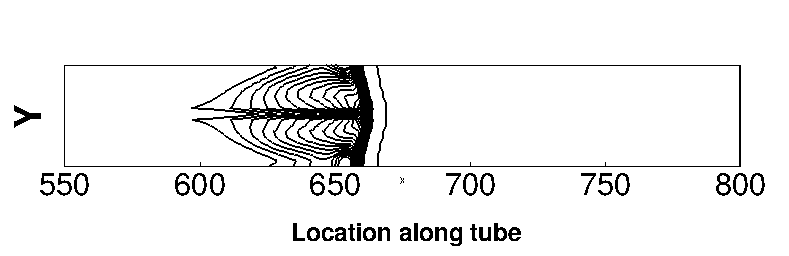}
		  \caption{HLLC}
		  \label{fig:hllc_movingshock}
		  \end{center}
		\end{subfigure}
		\begin{subfigure}[b]{0.5\textwidth}
		  \begin{center}
		  \includegraphics[width=6cm,height=2.5cm]{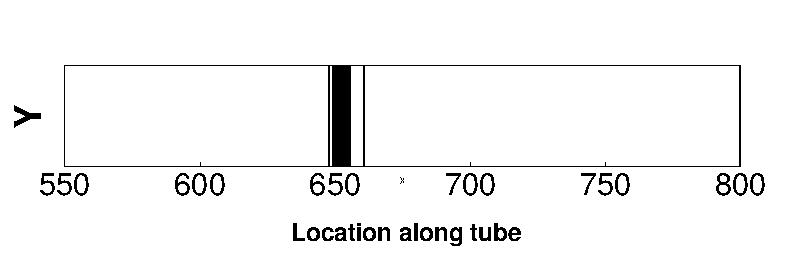}
		  \caption{HLLC-SWM-E}
		  \label{fig:hllcswm_e_movingshock}
		  \end{center}
		  \end{subfigure}
		\begin{subfigure}[b]{0.5\textwidth}
		  \begin{center}
		  \includegraphics[width=6cm,height=2.5cm]{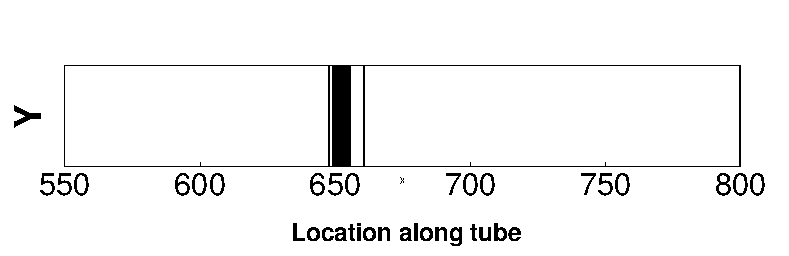}
		  \caption{HLLC-SWM-P}
		  \label{fig:hllcswm_p_movingshock}
		  \end{center}
		\end{subfigure}
	    \caption{Density contours for M=6 Quirk's moving shock instability problem.}
	   \label{fig:movingshockresults}
	    \end{figure}

Shock instability that occurs in a M=6  moving shock propagating down a computational tube filled a stationary fluid defined by $(\rho,u,v,p)=(1.4,1.0,0.0,1.0)$ 
was first reported by Quirk \cite{quirk1994}. 
The instability is triggered by perturbing the centerline grid of the computational domain to an order of 1E-6. First order solution is sought. 
The CFL for the computations were
taken to be 0.5 and simulations were run till shock reached a location x=650. The results showing 50 density
contours equally spanning values from 1.4 to 7.34 is shown in Fig.(\ref{fig:movingshockresults}).
%
\noindent Fig.(\ref{fig:hllc_movingshock}) clearly shows the deteriorated condition of the shock profile calculated by the original HLLC scheme. 
Figs.(\ref{fig:hllcswm_e_movingshock},\ref{fig:hllcswm_p_movingshock}) demonstrate that both the HLLC-SWM-E and the HLLC-SWM-P schemes are quite capable of 
suppressing the instabilities and maintaining a clean shock profile.

\subsection{Inclined stationary shock instability problem (Inviscid)}
\label{sec:inclinedshock_hllemfixed}
\begin{figure}[H]
		\begin{subfigure}[b]{0.3\textwidth}
		  \begin{center}
		 \includegraphics[width=4cm,height=4cm]{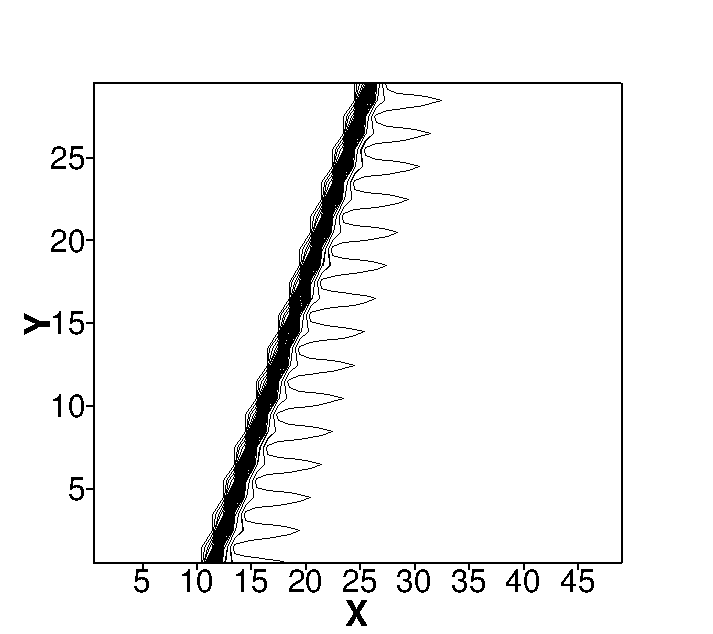}
		  \caption{HLLC}
		  \label{fig:hllc_inclinedshock}
		  \end{center}
		\end{subfigure}
		\begin{subfigure}[b]{0.3\textwidth}
		  \begin{center}
		  \includegraphics[width=4cm,height=4cm]{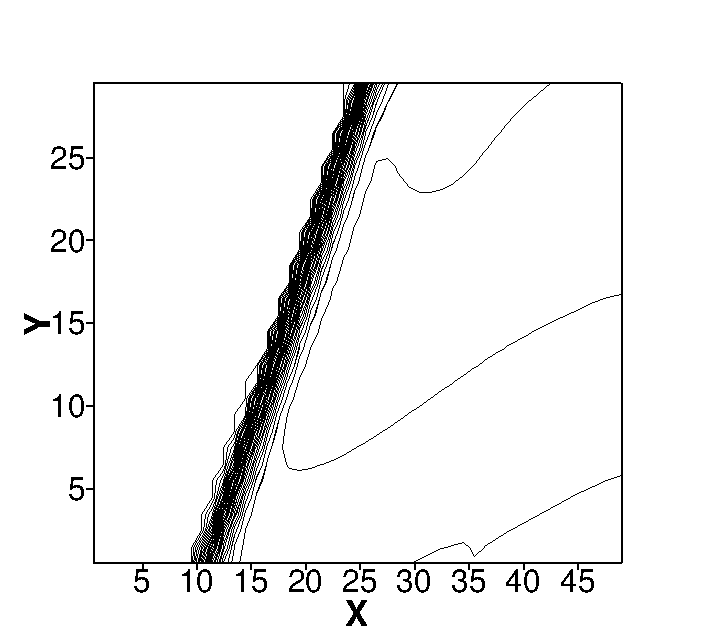}
		  \caption{HLLC-SWM-E}
		  \label{fig:hllc_swm_e_inclinedshock}
		  \end{center}
		  \end{subfigure}
		\begin{subfigure}[b]{0.3\textwidth}
		  \begin{center}
		  \includegraphics[width=4cm,height=4cm]{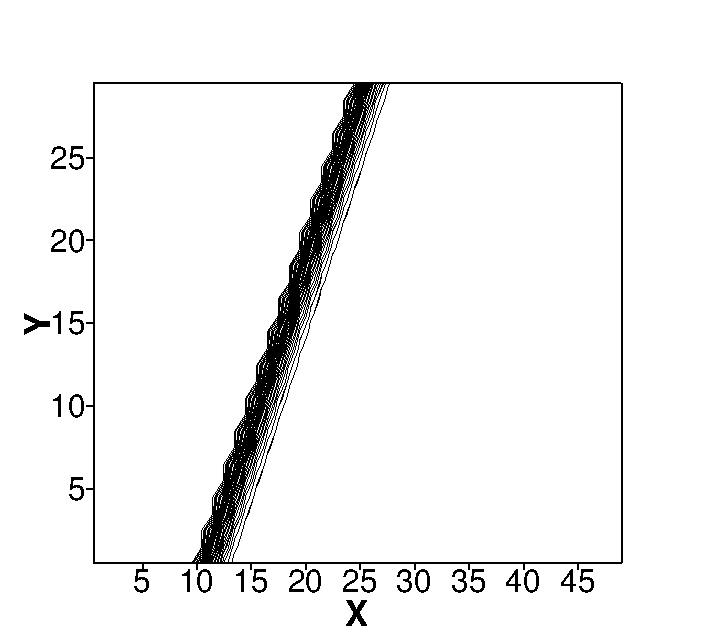}
		  \caption{HLLC-SWM-P}
		  \label{fig:hllc_swm_p_inclinedshock}
		  \end{center}
		\end{subfigure}
	   \caption{Density contours for M=7 stationary inclined shock problem.}
	   \label{fig:inclinedshock}
	    \end{figure}

This problem was reported in \cite{ohwada2013} and is a variant of the standing shock instability problem discussed previously. 
The initial shock wave of strength M=7 is set up along the line y=2(x-12) at an angle of $63.43^o$ with respect to the
global x-direction. The computation mesh consists of 50$\times$30 cells uniformly spanning a domain of size 50.0$\times$30.0. 
The left and right boundaries are maintained as supersonic inlet and subsonic outlet respectively while periodic conditions are applied to top and bottom walls.
CFL of the computation was chosen as 0.5 and first order solution was sought at t=5. 
Fig.(\ref{fig:inclinedshock}) shows the result of this experiment where thirty density contours 
uniformly spanning 1.0 to 5.44 are shown.

%
%

	    \begin{figure}[H]
		\begin{subfigure}[b]{0.3\textwidth}
		  \begin{center}
		 \includegraphics[width=4cm,height=4cm]{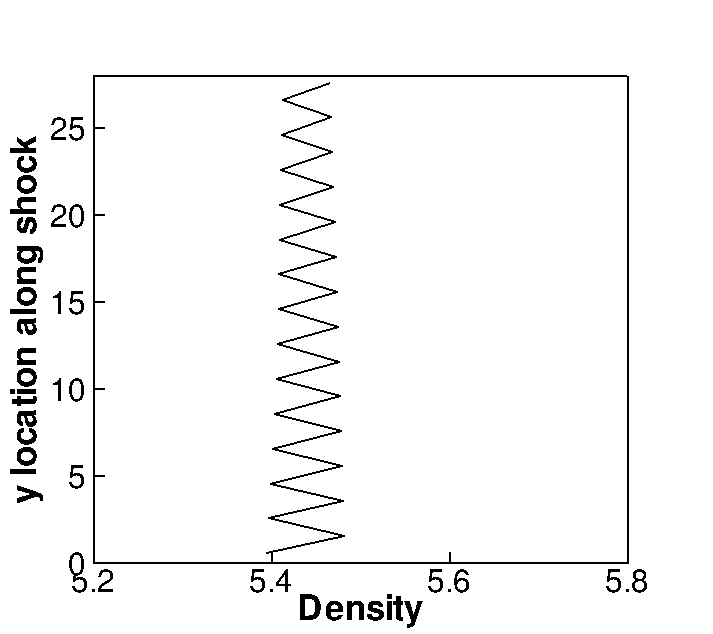}
		  \caption{HLLC}
		  \label{fig:hllc_inclinedshock_densityalongshock}
		  \end{center}
		\end{subfigure}
		\begin{subfigure}[b]{0.3\textwidth}
		  \begin{center}
		  \includegraphics[width=4cm,height=4cm]{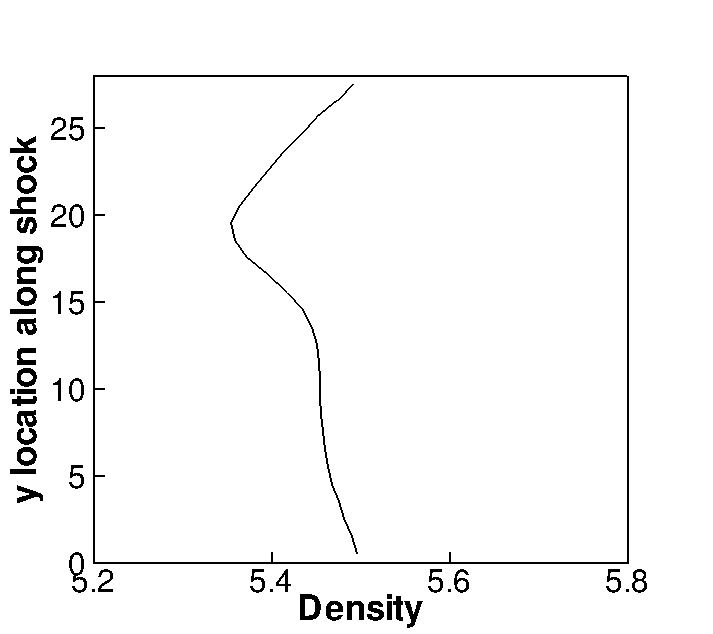}
		  \caption{HLLC-SWM-E}
		  \label{fig:hllc_swm_e_inclinedshock_densityalongshock}
		  \end{center}
		  \end{subfigure}
		\begin{subfigure}[b]{0.3\textwidth}
		  \begin{center}
		  \includegraphics[width=4cm,height=4cm]{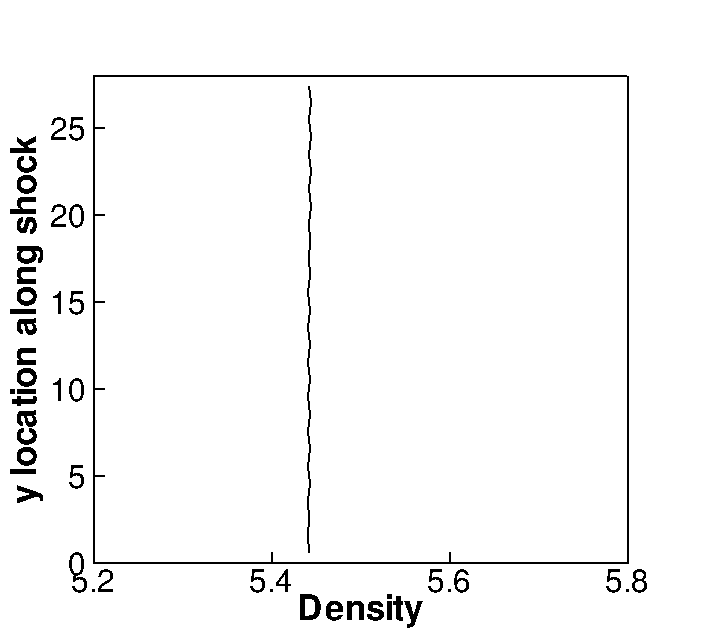}
		  \caption{HLLC-SWM-P}
		  \label{fig:hllc_swm_p_inclinedshock_densityalongshock}
		  \end{center}
		\end{subfigure}
	   \caption{Plot showing variation of density values in the immediate subsonic region along the shock profile for the M=7 stationary inclined shock problem in 
	   Sec.(\ref{sec:inclinedshock_hllemfixed}).}
	   \label{fig:inclinedshock_densityprofile}
	    \end{figure} 

Fig.(\ref{fig:hllc_inclinedshock}) shows the presence of instability in the HLLC scheme. Fig.(\ref{fig:hllc_inclinedshock_densityalongshock}) shows the density variation 
along the shock front extracted just behind the shock. The density variation has a typical saw tooth profile.
Fig.(\ref{fig:hllc_swm_e_inclinedshock}) shows the density contours computed by the HLLC-SWM-E scheme. It was noticed that although till t=2 units 
there was no trace of instability in the computed field, some sort of perturbations were observed in the simulation later. These perturbations however 
remained temporaly bounded and did not degrade the shock profile. The respective density variation is shown in Fig.(\ref{fig:hllc_swm_e_inclinedshock_densityalongshock}). 
It can be seen that the nature of these perturbations are not similar to the spatially oscillatory one observed for the HLLC scheme.
Fig.(\ref{fig:hllc_swm_p_inclinedshock}) shows the density contours computed by the HLLC-SWM-P scheme. The corresponding density variation is shown in 
Fig.(\ref{fig:hllc_swm_p_inclinedshock_densityalongshock}). It can be seen that introduction of the pressure sensor helps the HLLC-SWM-P scheme to outperform the HLLC-SWM-E scheme on 
this test case. The instability has been drastically minimized and the computed field is preserved very close to the exact solution.

\subsection{Supersonic flow over a blunt body (Inviscid)}

\begin{figure}[H]
		\begin{subfigure}[b]{0.3\textwidth}
		  \begin{center}
		 \includegraphics[width=2cm,height=5cm]{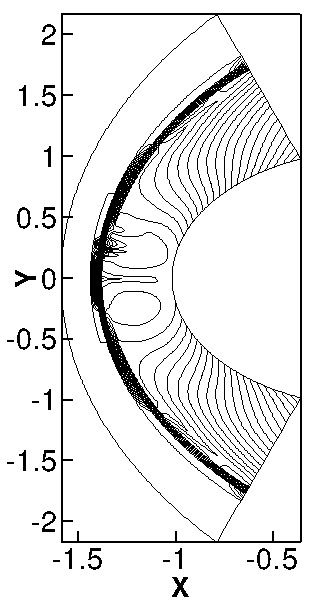}
		  \caption{HLLC}
		  \label{fig:hllc_carbuncle}
		  \end{center}
		\end{subfigure}
		\begin{subfigure}[b]{0.3\textwidth}
		  \begin{center}
		  \includegraphics[width=2cm,height=5cm]{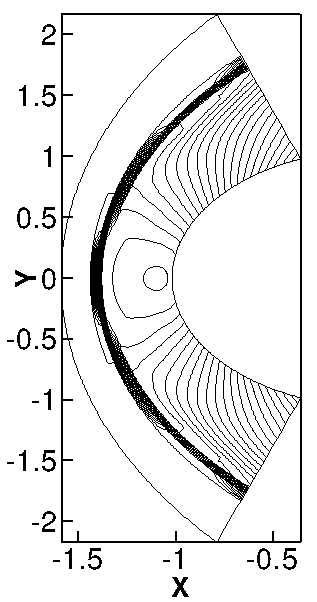}
		  \caption{HLLC-SWM-E}
		  \label{fig:hllcswm_e_carbuncle}
		  \end{center}
		  \end{subfigure}
		\begin{subfigure}[b]{0.3\textwidth}
		  \begin{center}
		  \includegraphics[width=2cm,height=5cm]{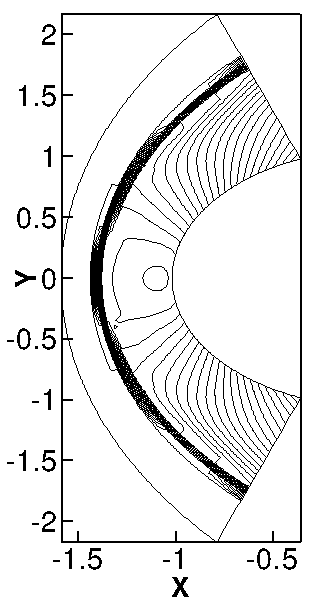}
		  \caption{HLLC-SWM-P}
		  \label{fig:hllcswm_p_carbuncle}
		  \end{center}
		\end{subfigure}
	  \caption{Density contours for M=20 supersonic flow over a cylindrical body.}
	   \label{fig:carbuncleresults_hllcswm}
	    \end{figure}

The steady state numerical solution of  hypersonic flow over cylindrical body is a routine test problem to investigate the susceptibility of numerical schemes
to a particularly dreaded form of shock instability: the carbuncle phenomenon. This failure is characterized by visible distortions that appear in the expected smooth bow shock profile in front of the body. 
In this work, a cylindrical body of unit radius is subjected to a M=20 flow. We use 320$\times$20 body fitted structured quadrilateral cells in circumferential and 
radial directions respectively. Detailed problem description can be found in \cite{huang2011}.
The computation of this problem is carried out with plain first order accuracy. 
The CFL for the computations were taken to be 0.5 and simulations were run for 
30,000 iterations. The results showing twenty density contours equally spanning value from 1.4 to 8.5 is shown in Fig.(\ref{fig:carbuncleresults_hllcswm}). 



\begin{figure}
		\begin{subfigure}[b]{0.5\textwidth}
		  \begin{center}
		 \includegraphics[width=5cm,height=5cm]{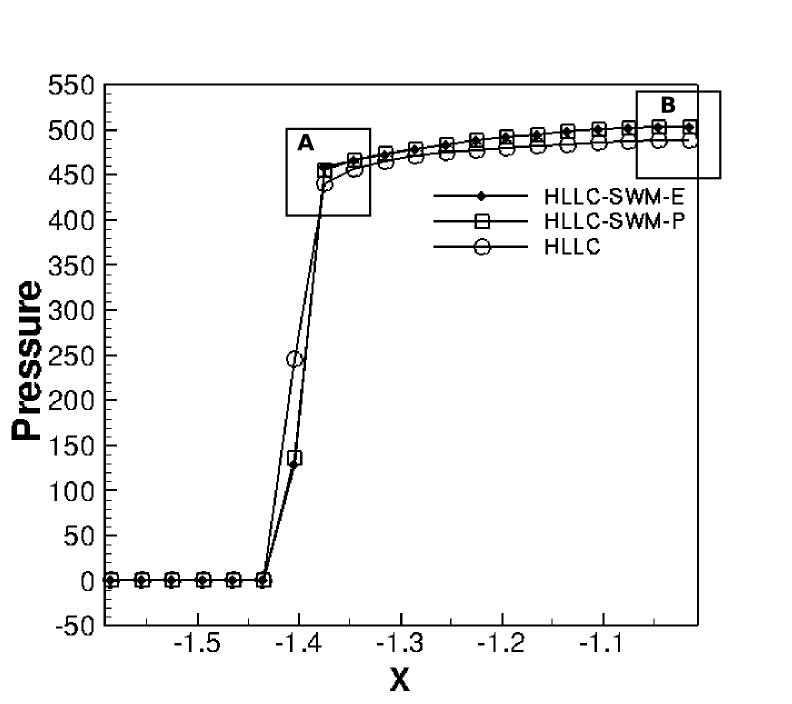}
		  \caption{Centerline pressures}
		  \label{fig:centerline_comparison_overall}
		  \end{center}
		\end{subfigure}
		\begin{subfigure}[b]{0.5\textwidth}
		  \begin{center}
		  \includegraphics[width=5cm,height=5cm]{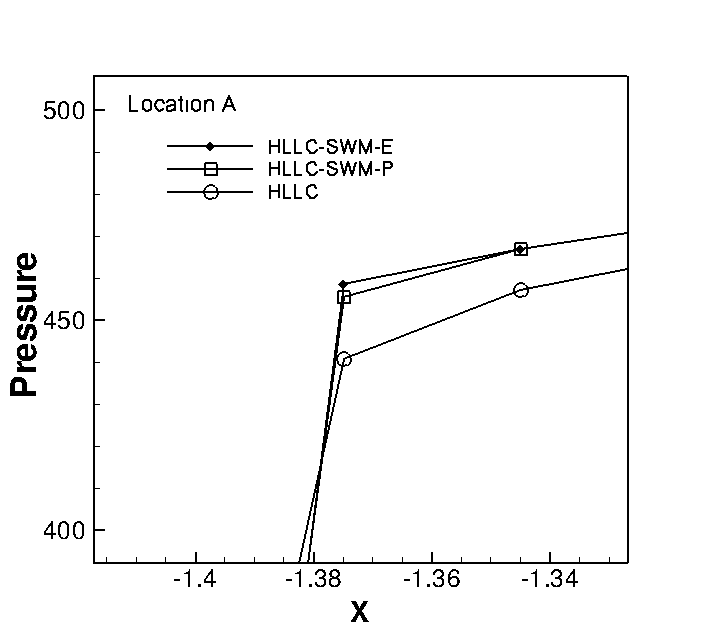}
		  \caption{Location A}
		  \label{fig:centerline_comparison_A}
		  \end{center}
		  \end{subfigure}
		\begin{subfigure}[b]{0.5\textwidth}
		  \begin{center}
		  \includegraphics[width=5cm,height=5cm]{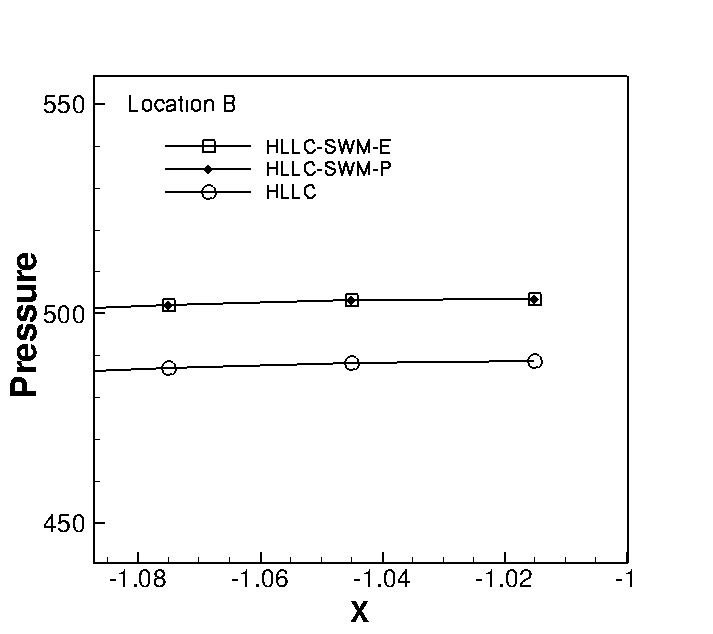}
		  \caption{Location B}
		  \label{fig:centerline_comparison_B}
		  \end{center}
		\end{subfigure}
	  \caption{Comparison of centerline pressures between HLLC and the proposed HLLC-SWM schemes.}
	   \label{fig:centerline_comparison}
	    \end{figure}

\noindent Fig.(\ref{fig:hllc_carbuncle}) clearly shows the onset of instabilities in the shock profile calculated by the HLLC scheme. Figs.(\ref{fig:hllcswm_e_carbuncle},
\ref{fig:hllcswm_p_carbuncle})
on the other hand demonstrates that the HLLC-SWM schemes are quite capable of computing clean shock profiles and post shock subsonic reions which are 
completely free of any instabilities. 
A quantitative review of the performance of these schemes is given in Fig.(\ref{fig:centerline_comparison_overall}) which shows 
the pressure value extracted along the stagnation line (j=160). Specific locations of interest on the plot are labeled as  \textquoteleft A \textquoteright and
\textquoteleft B \textquoteright and are shown magnified in Figs.(\ref{fig:centerline_comparison_A}) and (\ref{fig:centerline_comparison_B}).

	    
\noindent Referring to Fig.(\ref{fig:centerline_comparison_A}), at location \textquoteleft A \textquoteright, it is seen that the instability has rendered a much diffused 
shock in case of the HLLC scheme. In comparison, while both HLLC-SWM schemes produce crisper shocks, the HLLC-SWM-E scheme seems to be slightly better than HLLC-SWM-P scheme in predicting
the after shock pressure value. Location 
\textquoteleft B \textquoteright which is close to the stagnation point clearly shows that both the HLLC-SWM schemes produce much accurate stagnation pressure values 
(which is closer to the expected inviscid value of 515.60) as compared to the HLLC scheme. 
This example shows the fidelty of the proposed HLLC-SWM schemes in computing Carbuncle free solutions on such severe grids.

\subsection{Double Mach Reflection problem (Inviscid)}
\begin{figure}[!]
		\begin{subfigure}[b]{0.5\textwidth}
		  \begin{center}
		 \includegraphics[width=6.5cm,height=3.5cm]{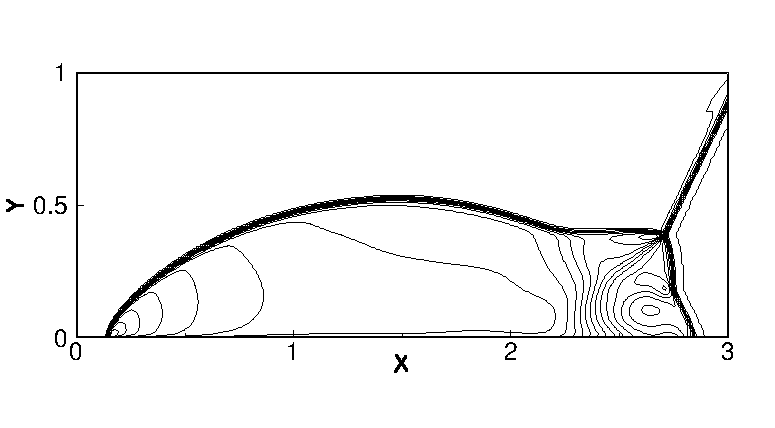}
		  \caption{HLLC}
		  \label{fig:hllc_dmr}
		  \end{center}
		\end{subfigure}
		\begin{subfigure}[b]{0.5\textwidth}
		  \begin{center}
		  \includegraphics[width=6.5cm,height=3.5cm]{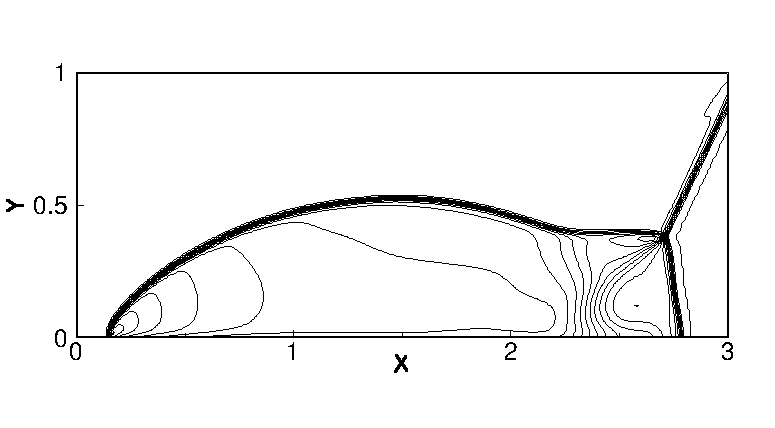}
		  \caption{HLLC-SWM-E}
		  \label{fig:hllcswm_e_dmr}
		  \end{center}
		  \end{subfigure}
		\begin{subfigure}[b]{0.5\textwidth}
		  \begin{center}
		  \includegraphics[width=6.5cm,height=3.5cm]{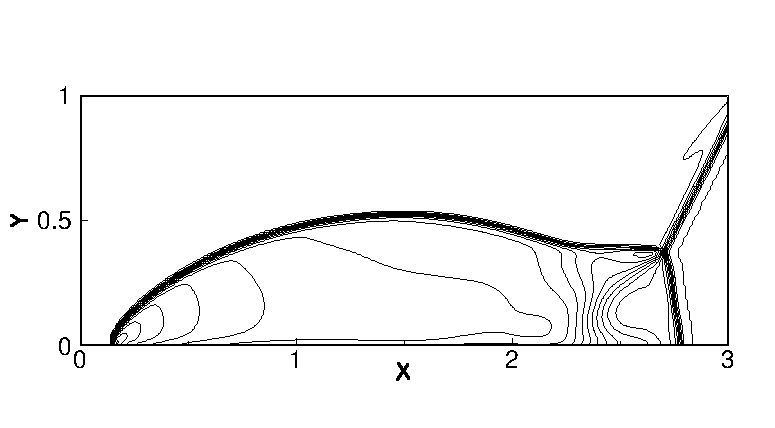}
		  \caption{HLLC-SWM-P}
		  \label{fig:hllcswm_p_dmr}
		  \end{center}
		\end{subfigure}
	  \caption{Density contours for double Mach reflection problem.}
	   \label{fig:dmrresults_hllcswm}
	    \end{figure}

Development of kinked Mach stem in a situation of a double Mach reflection has been reported in various literatures \cite{quirk1994, woodward1984}. 
An oblique shock corresponding to $M=10$, making a $60^o$ angle with the bottom wall is made to propagate through a domain of size
$4.0\times1.0$ divided into $480\times120$ structured Cartesian cells containing stationary fluid. Detailed problem description can be found in \cite{woodward1984}.
First order solution in time and space is sought for this case.
The simulation is run till 
$t=0.02$ with CFL of 0.8. Fig.(\ref{fig:dmrresults_hllcswm}) shows results of the experiment with twenty five equispaced density contours spanning values from 1.4 to 21.

\noindent In Fig.(\ref{fig:hllc_dmr}), the presence of kinked Mach stem and the subsequent triple point produced by the HLLC scheme is clearly visible. However, both the
HLLC-SWM schemes produce clean shock profiles which are devoid of the kink and triple point. 

\subsection{Diffraction of a moving normal shock over a 90$^{0}$ corner (Inviscid)}
\begin{figure}[!]
		\begin{subfigure}[b]{0.5\textwidth}
		  \begin{center}
		 \includegraphics[width=7cm,height=4cm]{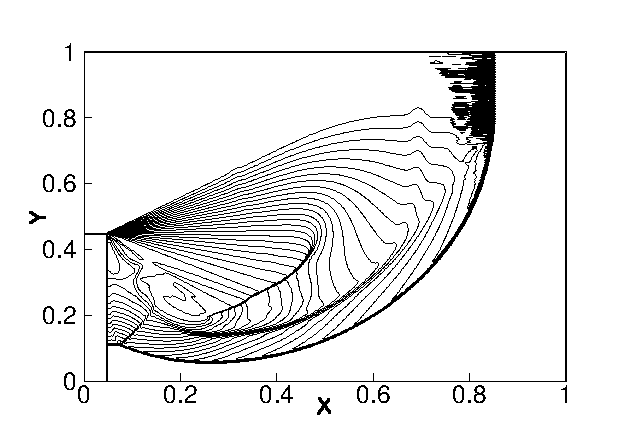}
		  \caption{HLLC}
		  \label{fig:hllc_supersoniccorner}
		  \end{center}
		\end{subfigure}
		\begin{subfigure}[b]{0.5\textwidth}
		  \begin{center}
		  \includegraphics[width=7cm,height=4cm]{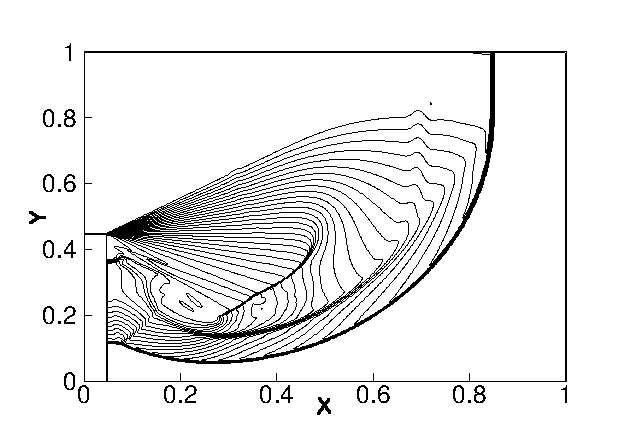}
		  \caption{HLLC-SWM-E}
		  \label{fig:hllcswm_e_supersoniccorner}
		  \end{center}
		  \end{subfigure}
		\begin{subfigure}[b]{0.5\textwidth}
		  \begin{center}
		  \includegraphics[width=7cm,height=4cm]{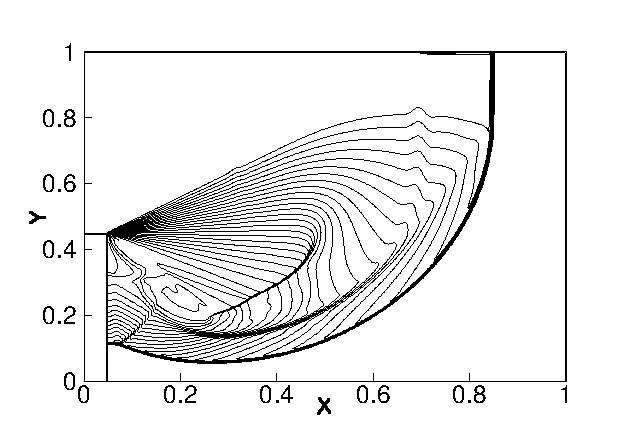}
		  \caption{HLLC-SWM-P}
		  \label{fig:hllcswm_p_supersoniccorner}
		  \end{center}
		\end{subfigure}
	 \caption{Density contours for M=5.09 normal shock diffraction around a $90^o$ corner.}
	   \label{fig:supersoniccornerresults}
	    \end{figure}

The problem of a Mach 5.09 normal shock's sudden expansion around a $90^o$ corner was studied in \cite{quirk1994}. We use an unit dimensional domain 
meshed with 400 $\times$ 400 cells with the right angled corner where the initial normal shock stands is located at $x=0.05$, $y=0.45$. 
The problem is computed to second order accuracy. Simulations are run for $t=0.1561$ units with CFL of 
0.4. Thirty density contours equally spanning values of 0 to 7 are
shown in Fig.(\ref{fig:supersoniccornerresults}).

\noindent Fig.(\ref{fig:hllc_supersoniccorner}) clearly demonstrates the extent of instability present in the computed shock profile produced by the HLLC 
scheme. A part of the normal shock in the right corner is completely distorted in this case. However, both the HLLC-SWM schemes are able to satisfactorily compute
the solution without producing any trace of instability in the normal shock. It must also be noted that HLLC-SWM schemes
does not produce any unphysical expansion shocks like the Roe scheme \cite{quirk1994}.
 
\subsection{ Supersonic flow over forward facing step (Inviscid)}

\begin{figure}[!]
		\begin{subfigure}[b]{0.5\textwidth}
		  \begin{center}
		 \includegraphics[width=6.5cm,height=3.5cm]{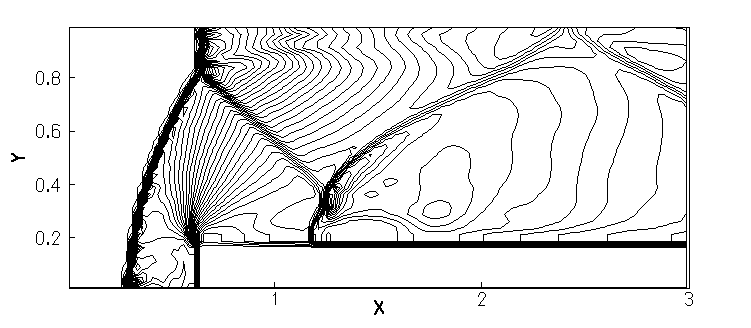}
		  \caption{HLLC}
		  \label{fig:hllc_ffs}
		  \end{center}
		\end{subfigure}
		\begin{subfigure}[b]{0.5\textwidth}
		  \begin{center}
		  \includegraphics[width=6.5cm,height=3.5cm]{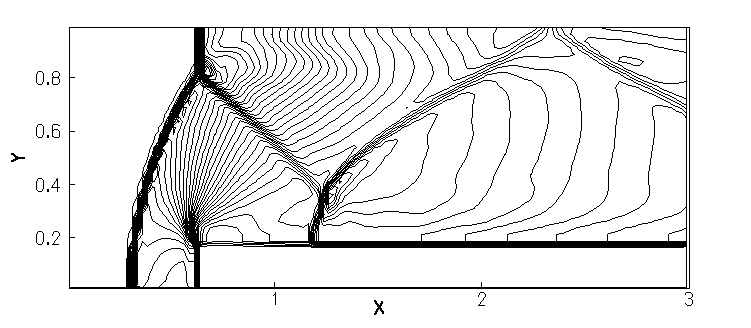}
		  \caption{HLLC-SWM-E}
		  \label{fig:hllcswm_e_ffs}
		  \end{center}
		  \end{subfigure}
		\begin{subfigure}[b]{0.5\textwidth}
		  \begin{center}
		  \includegraphics[width=6.5cm,height=3.5cm]{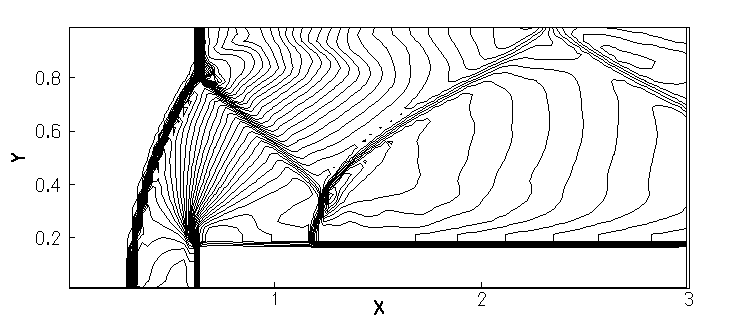}
		  \caption{HLLC-SWM-P}
		  \label{fig:hllcswm_p_ffs}
		  \end{center}
		\end{subfigure}
	\caption{Density contours for M=3 flow over a forward facing step problem.}
	   \label{fig:ffsresults}
	    \end{figure}

A M=3 supersonic flow over forward facing step is another standard problem that has been extensively studied earlier in 
\cite{woodward1984}. The computational domain is of size $3\times1$ and is meshed with $120\times40$ structured Cartesian cells. A 0.2 units high step is located at a distance of 0.6 units from 
the inlet. The problem is computed to second order accuracy. The simulations are run for t=4 with CFL as 0.5. 
Fig.(\ref{fig:ffsresults}) shows forty equispaced density contours spanning 0.2 to 7.0. 

\noindent It is seen from Fig.(\ref{fig:hllc_ffs}) that predominant oscillations are visible in the shock profile computed by HLLC scheme. These are most
prominent at the normal shock stems of the primary shock near the bottom and top boundaries. Further, HLLC solution also has a severe kinked stem in the reflected shock. 
In comparison, solution computed by both HLLC-SWM scheme shown in Figs.(\ref{fig:hllcswm_e_ffs},\ref{fig:hllcswm_p_ffs}) shows clean shock profiles with no 
remnants of instabilities visible. The infamous kink has also improved considerably in these solutions.

\subsection{Two dimensional shear flow (Inviscid)}
\label{sec:inviscidshearflow}
\begin{figure}[H]
		\begin{subfigure}[b]{0.3\textwidth}
		  \begin{center}
		 \includegraphics[width=4cm,height=4cm]{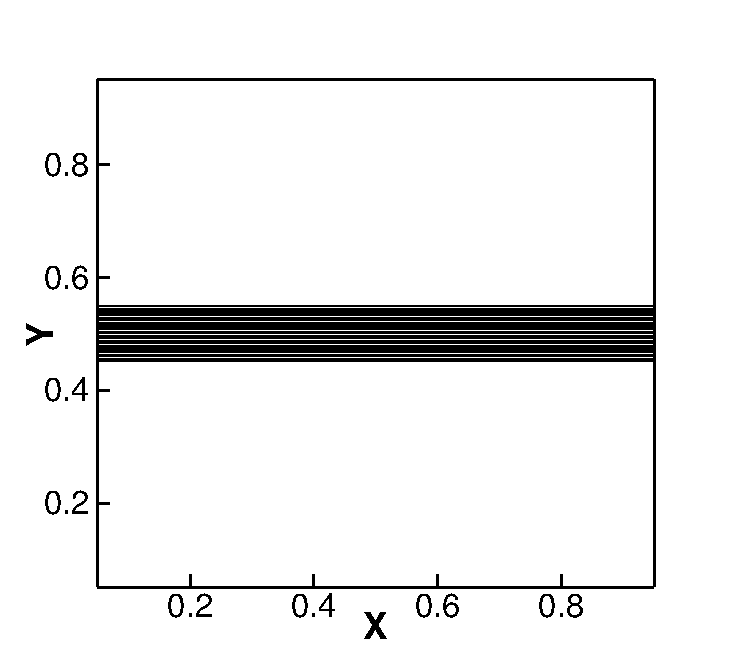}
		  \caption{HLLC}
		  \label{fig:hllc_shear}
		  \end{center}
		\end{subfigure}
		\begin{subfigure}[b]{0.3\textwidth}
		  \begin{center}
		  \includegraphics[width=4cm,height=4cm]{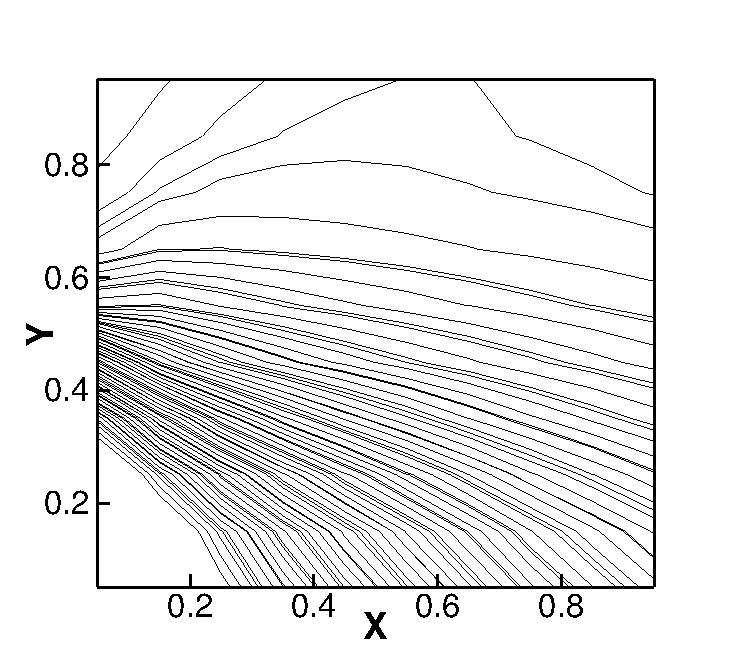}
		  \caption{HLLE}
		  \label{fig:hll_shear}
		  \end{center}
		  \end{subfigure}
		\begin{subfigure}[b]{0.3\textwidth}
		  \begin{center}
		  \includegraphics[width=4cm,height=4cm]{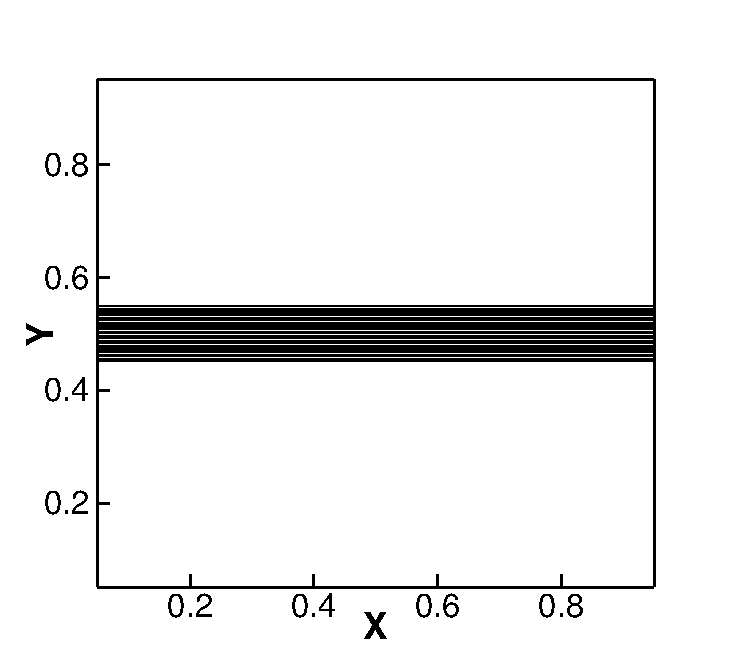}
		  \caption{HLLC-SWM-E}
		  \label{fig:hllcswm_e_shear}
		  \end{center}
		  \end{subfigure}
		\begin{subfigure}[b]{0.3\textwidth}
		  \begin{center}
		  \includegraphics[width=4cm,height=4cm]{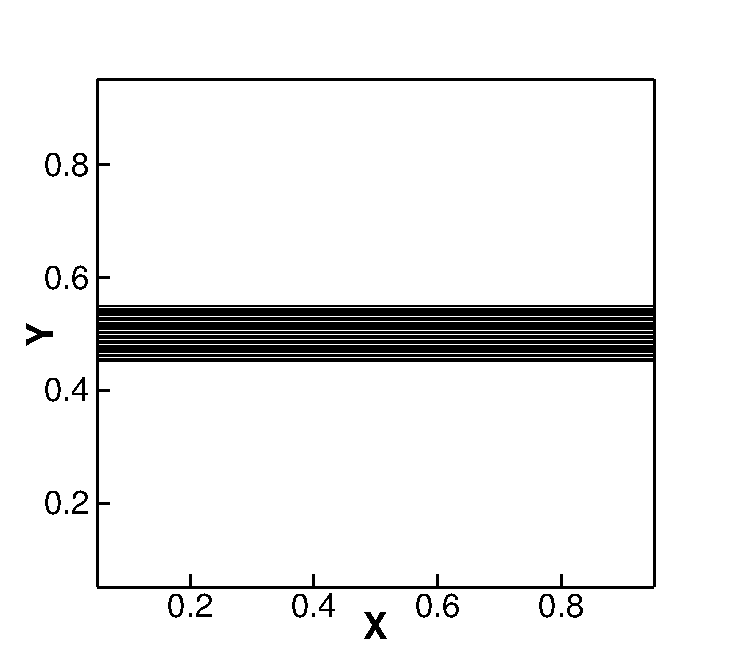}
		  \caption{HLLC-SWM-P}
		  \label{fig:hllcswm_p_shear}
		  \end{center}
		\end{subfigure}
		\begin{subfigure}[b]{0.5\textwidth}
		  \begin{center}
		  \includegraphics[width=4cm,height=4cm]{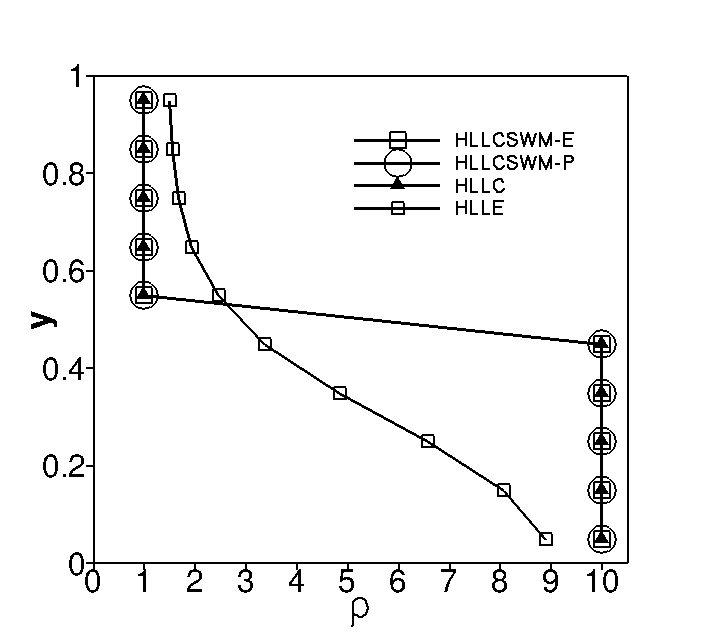}
		  \caption{HLLC-SWM-P}
		  \label{fig:shearresults_densityvariation_hllcswm}
		  \end{center}
		\end{subfigure}
	\caption{Figures (a), (b), (c) and (d) show the density contours for the two dimensional supersonic shear flow while Figure (e) shows the density variation along 
	y-direction at the center of the domain.}
	   \label{fig:shearresults_hllcswm}
	    \end{figure}

The problem describes two fluids with different densities sliding at different speeds over each other and investigates the inviscid contact capturing ability of a given scheme
\cite{wada1997}.
The conditions for the top and bottom fluids are chosen as $(\rho, p,M)_{top}$ = (1, 1, 2) and $(\rho, p,M)_{bottom}$ = (10, 1, 1.1) respectively.
A coarse grid consisting of 10$\times$10 cells on a domain of 1.0$\times$1.0 is used. CFL for the simulations are taken to be 1.0 and simulations were run for 1000 iterations. 
All simulations are plain
first order accurate.
The results showing fifty equispaced density contours spanning 1.0 to 10.0 is shown in Fig.(\ref{fig:shearresults_hllcswm}). To clearly distinguish between the 
behaviours of contact preserving and contact dissipative schemes, the solution for the HLLE scheme is also given for comparison. 


\noindent It can be seen from Fig.(\ref{fig:shearresults_hllcswm}) that behaviours of the HLLC-SWM schemes closely matches that of the HLLC scheme. All these schemes are 
able to clearly preserve the inviscid contact. On the other hand, the HLLE scheme produces a very diffused density profile.
Fig.(\ref{fig:shearresults_densityvariation_hllcswm}) shows variation of density along y-location at the center of the domain which confirms the observation made above.
From the plot it can be inferred 
that the modified dissipation vector $\overline{\mathbf{D}}_{HLL}$ that introduces additional dissipation into the HLLC scheme to deal with shock instabilities
has not affected the
ability to resolve inviscid contact discontinuities.

	    
%

\subsection{Laminar flow over flat plate (Viscous)}
\label{sec:flatplate}
\begin{figure}
	    \centering
	    \includegraphics[scale=0.35]{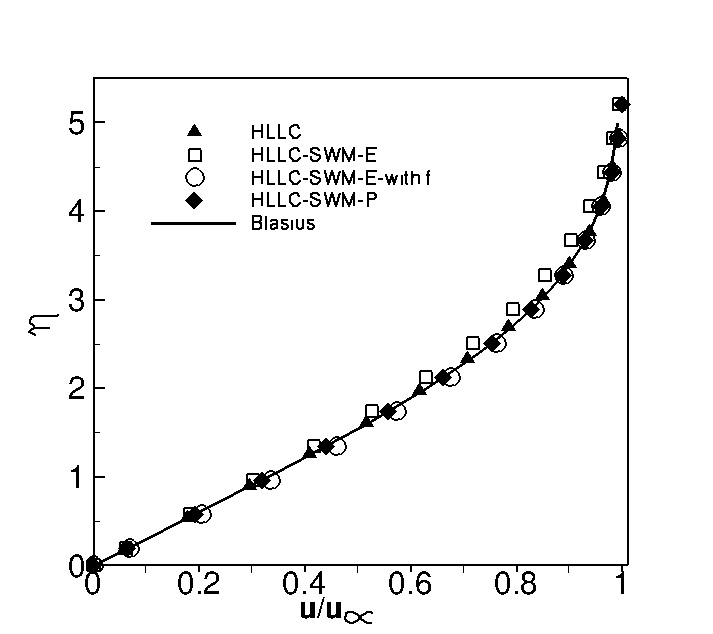}
	   \caption{Result for laminar flow over a flat plate.}
	   \label{fig:laminarflatplate}
	    \end{figure}

The original HLLC scheme is well known for its ability to 
accurately resolve shear dominated viscous flow features including practically significant boundary layer flows. It is hence interesting to investigate 
how modified dissipation vector $\overline{\mathbf{D}}_{HLL}$ will affect the performance of the HLLC scheme on these types of flows. To perform this study, 
the test case of a $M = 0.1$ subsonic laminar flow of air over a flat plate of length L=0.3048 m in atmospheric conditions is chosen. The total length of the domain 
is 0.381 m in x direction and 0.1 m in y direction. The domain is divided into 31$\times$33 Cartesian cells  with 15 cells within the boundary layer. 
Viscous fluxes were discretized using simple averaging. 
CFL was taken to be 0.7. 
The flow was considered to have achieved steady state when the horizontal velocity residuals dropped to the order of $1e^{-7}$. 
The normalized longitudinal velocity profiles $(\frac{u}{u_{\infty}})$ are plotted against the Blasius parameter $\eta=y\sqrt{u_{\infty}/\mu L}$ in Fig.(\ref{fig:laminarflatplate}).  

\noindent It is seen from Fig.(\ref{fig:laminarflatplate}) that the HLLC-SWM-E scheme suffers a slight loss of accuracy within the boundary layer. 
However close to the wall, the HLLC-SWM-E scheme is as accurate as the HLLC scheme in resolving the velocity gradients. 
On the other hand, the HLLC-SWM-P scheme is remarkably accurate everywhere within the boundary layer and exactly matches the Blasius profile. 
As discussed previously, the dissipative behavior of the HLLC-SWM-E scheme could be explained from the fact that since the $\epsilon^{L,R}$ in Eq.(\ref{eqn:hllc_modifiedwavespeedestimates}) 
computes the maximum difference 
in wavespeeds at any interface, it could be assuming a non zero value in the shear layer due to existence of gradients $\Delta_x (u,v)$ and $\Delta_y (u,v)$.
This would then enlarge the magnitude of the $\overline{\mathbf{D}}_{HLL}$ term thus making the HLLC-SWM-E scheme more dissipative.
To control this unwanted dissipation in the shear dominated
low Mach number flows, a simple and inexpensive Mach number based switching function $f$ is defined as, 
\begin{align}
\begin{cases}
    f = sin( (\pi/2) * (M/0.3)),& \text{if } M\;\; \leq\;\;0.3 \\
    f = 1                      ,& \text{if } M\;\; >\;\; 0.3
\end{cases}
\label{eqn:switchingfunction}
\end{align}
where $M$ is the local flow Mach number. The function $f$ can now be integrated into Eq.(\ref{eqn:hllc_modifiedwavespeedestimates}) as,
\begin{align}
\nonumber
 \overline{S_L} = S_L - f\alpha\epsilon\\
 \overline{S_R} = S_R + f\alpha\epsilon
 \label{eqn:hllc_modifiedwavespeedestimates_withswitchingfunction}
\end{align}
It is apparent that the switching function $f$ smoothly transits from its value of 1 for $M>0.3$ to its value of 0 as $M\rightarrow$ 0. 
This enables reduction of dissipation in the HLLC-SWM-E scheme thus recovering the full accuracy of the original HLLC scheme within the viscous layers.
The results showing the effect of $f$ on HLLC-SWM-E is also shown in Fig.(\ref{fig:laminarflatplate}). It is evident that 
introduction of switching function clearly restores the accuracy of the solution.  

\section{Concluding remarks}
\label{sec:conclusions}

We have proposed a simple cure for numerical shock instabilities in the Harten\-Lax\-vanLeer with contact (HLLC) approximate Riemann solver.
Firstly we showed that the HLLC scheme can be split into its respective diffusive HLL component and an antidiffusive component responsible for accuracy on 
contact and shear waves. Once identified, we suggest a simple modificiation to certain nonlinear wavespeed estimates appearing only in the numerical dissipation 
of the HLL component that increases its magnitude. Two entropy fix based strategies for obtaining multidimensional dissipation in the vicinity of a shock wave was presented which resulted in 
the HLLC-SWM-E and HLLC-SWM-P variants of the proposed framework. 
Through a linear analysis of a mean flow with perturbations and a matrix based stability analysis of an isolated two dimensional stationary normal shock, the theoretical 
effectiveness of these schemes in computing shock stable solutions over a wide range of Mach numbers was established.
Further, using numerical test cases we show that the HLLC-SWM schemes provide instability free, clean shock profiles on a variety of problems while retaining the 
positivity preserving and exact inviscid contact abilities of the underlying HLLC scheme.
When computing the problem of a subsonic laminar flow over flat plate, it was found that the HLLC-SWM-P scheme was as accurate as the HLLC scheme while the HLLC-SWM-E scheme 
introduced slight
extraneous dissipation into the solution. A simple strategy that involves a Mach number based switching function is proposed to recover the accuracy of the HLLC-SWM-E scheme
on such problems. The success of these HLLC-SWM schemes in dealing with shock instability phenomenon alludes to the fact that cures for these spurious failures can be 
designed by introducing 
adequate multidimensional dissipation in the vicinity of a shock front. Lastly, the development of the HLLC-SWM-P scheme suggests that it is possible to circumvent
the conjecture in Gressier et al \cite{gressier2000} and create a robust shock 
stable upwind scheme that is accurate on contact and shear waves.

%
%

\bibliographystyle{unsrtnat}
\bibliography{References}

\begin{thebibliography}{45}
\providecommand{\natexlab}[1]{#1}
\providecommand{\url}[1]{\texttt{#1}}
\expandafter\ifx\csname urlstyle\endcsname\relax
  \providecommand{\doi}[1]{doi: #1}\else
  \providecommand{\doi}{doi: \begingroup \urlstyle{rm}\Url}\fi

\bibitem[Roe(1981)]{roe1981}
P.L. Roe.
\newblock {Approximate Riemann solvers, parameter vectors, and difference
  schemes}.
\newblock \emph{J. Comput. Phys.}, 43:\penalty0 357--372, 1981.

\bibitem[Einfeldt et~al.(1991)Einfeldt, Munz, Roe, and Sjogreen]{einfeldt1991}
B.~Einfeldt, C.~D. Munz, P.~L. Roe, and B.~Sjogreen.
\newblock {On Gudonov type methods near low densities}.
\newblock \emph{J. Comput. Phys.}, 92:\penalty0 273--295, 1991.

\bibitem[Toro et~al.(1994)Toro, Spruce, and Speares]{toro1994}
E.~F. Toro, M.~Spruce, and W.~Speares.
\newblock {Restoration of the contact surface in the HLL-Riemann Solver}.
\newblock \emph{Shock Waves}, 4:\penalty0 25--34, 1994.

\bibitem[Harten et~al.(1983)Harten, Lax, and van Leer]{hll1983}
A.~Harten, P.~D. Lax, and B.~van Leer.
\newblock {On upstream differencing and Godunov-type schemes for hyperbolic
  conservation laws}.
\newblock \emph{SIAM Rev.}, 25:\penalty0 35--61, 1983.

\bibitem[Einfeldt(1988)]{einfeldt1988}
Bernd Einfeldt.
\newblock On godunov-type methods for gas dynamics.
\newblock \emph{SIAM Journal on Numerical Analysis}, 25\penalty0 (2):\penalty0
  294--318, 1988.

\bibitem[Henderson and Menart(2007)]{henderson2007}
S.~Henderson and J.~Menart.
\newblock {Grid Study on Blunt Bodies with the Carbuncle Phenomenon}.
\newblock \emph{39th AIAA Thermophysics conference, Miami, Florida}, 2007.

\bibitem[Dumbser et~al.(2004)Dumbser, Moschetta, and Gressier]{dumbser2004}
M.~Dumbser, J.~M. Moschetta, and J.~Gressier.
\newblock {A matrix stability analysis of the carbuncle phenomenon}.
\newblock \emph{J. Comput. Phys.}, 197:\penalty0 647--670, 2004.

\bibitem[Chauvat et~al.(2005)Chauvat, Moschetta, and Gressier]{chauvat2005}
Y.~Chauvat, J.~M. Moschetta, and J.~Gressier.
\newblock {Shock wave numerical structure and the carbuncle phenomenon}.
\newblock \emph{Int. J. Numer. Methods Fluids}, 47:\penalty0 903--909, 2005.

\bibitem[Gressiera and Moschetta(2000)]{gressier2000}
J.~Gressiera and J.-M. Moschetta.
\newblock {Robustness versus accuracy in shock-wave computations}.
\newblock \emph{Int. J. Numer. Methods Fluids}, 33:\penalty0 313--332, 2000.

\bibitem[Quirk(1994)]{quirk1994}
J.~J. Quirk.
\newblock {A contribution to the great Riemann solver debate}.
\newblock \emph{Int. J. Numer. Methods Fluids}, 18:\penalty0 555--574, 1994.

\bibitem[Pandolfi and D’Ambrosio(2001)]{pandolfi2001}
M.~Pandolfi and D.~D’Ambrosio.
\newblock {Numerical instabilities in upwind methods: Analysis and cures for
  the “Carbuncle” phenomenon}.
\newblock \emph{J. Comput. Phys.}, 166:\penalty0 271--301, 2001.

\bibitem[Xu and Li(2001)]{xu2001}
Kun Xu and Zuowu Li.
\newblock {Dissipative mechanism in Godunov-type schemes}.
\newblock \emph{Int. J. Numer. Methods Fluids}, 37\penalty0 (1):\penalty0
  1--22, 2001.

\bibitem[Xie et~al.(2017)Xie, Li, Li, Tian, and Pan]{xie2017}
W.~Xie, W.~Li, H.~Li, Z.~Tian, and S.~Pan.
\newblock {On numerical instabilities of Godunov-type schemes for strong
  shocks}.
\newblock \emph{Journal of Computational Physics}, 350:\penalty0 607--637,
  2017.

\bibitem[Ismail(2006)]{farhad2006}
F.~Ismail.
\newblock \emph{Toward A Reliable Prediction Of Shocks In Hypersonic Flow:
  Resolving Carbuncles With Entropy And Vorticity Control}.
\newblock PhD thesis, University of Michigan, 2006.

\bibitem[Shen et~al.(2016)Shen, Yan, and Yuan]{shen2016}
Z.~Shen, W.~Yan, and G.~Yuan.
\newblock A robust hllc-type riemann solver for strong shock.
\newblock \emph{J. Comput. Phys}, 309:\penalty0 185 -- 206, 2016.

\bibitem[Ren(2003)]{ren2003}
Y.-X. Ren.
\newblock A robust shock-capturing scheme based on rotated riemann solvers.
\newblock \emph{Comput and Fluids}, 32:\penalty0 1379 -- 1403, 2003.

\bibitem[Sanders et~al.(1998)Sanders, Morano, and Druguet]{sanders1998}
R.~Sanders, E.~Morano, and M.~C. Druguet.
\newblock {Multidimensional dissipation for upwind schemes:Stability and
  applications to gas dynamics}.
\newblock \emph{J. Comput. Phys.}, 145:\penalty0 511--537, 1998.

\bibitem[Shen et~al.(2014)Shen, Yan, and Yuan]{shen2014}
Z.~Shen, W.~Yan, and G.~Yuan.
\newblock {A stability analysis of Hybrid schemes to cure shock instability}.
\newblock \emph{Commun. Comput. Phys.}, 15:\penalty0 1320--1342, 2014.

\bibitem[Wu et~al.(2010)Wu, Shen, and Shen]{wu2010}
H.~Wu, L.~Shen, and Z.~Shen.
\newblock {A hybrid numerical method to cure numerical shock instability}.
\newblock \emph{Commun. Comput. Phys.}, 8:\penalty0 1264--1271, 2010.

\bibitem[Wang et~al.(2016)Wang, Deng, Wang, and Dong]{dongfang2016}
Dongfang Wang, Xiaogang Deng, Guangxue Wang, and Yidao Dong.
\newblock Developing a hybrid flux function suitable for hypersonic flow
  simulation with high-order methods.
\newblock \emph{Int. J. Numer. Methods Fluids}, 81\penalty0 (5):\penalty0
  309--327, 2016.

\bibitem[Nishikawa and Kitamura(2008)]{nishikawa2008}
H.~Nishikawa and K.~Kitamura.
\newblock Very simple carbuncle free boundary layer resolving rotated hybrid
  riemann solvers.
\newblock \emph{J. Comput. Phys}, 227:\penalty0 2560--2581, 2008.

\bibitem[Huang et~al.(2011)Huang, Wu, Yu, and Yan]{huang2011}
K.~Huang, H.~Wu, H.~Yu, and D.~Yan.
\newblock Cures for numerical shock instability in hllc solver.
\newblock \emph{Int. J. Numer. Methods Fluids}, 65:\penalty0 1026--1038, 2011.

\bibitem[Kim et~al.(2009)Kim, Lee, Lee, and Jeung]{kim2009}
S.D. Kim, B.J. Lee, H.J. Lee, and I.-S Jeung.
\newblock Robust hllc riemann solver with weighted average flux scheme for
  strong shock.
\newblock \emph{J. Comput. Phys}, 228:\penalty0 7634 -- 7642, 2009.

\bibitem[Kim et~al.(2010)Kim, Lee, Lee, Jeung, and Choi]{kim2010}
S.~D. Kim, B.~J. Lee, H.~J. Lee, I.-S. Jeung, and J.-Y Choi.
\newblock Realization of contact resolving approximate riemann solvers for
  strong shock and expansion flows.
\newblock \emph{Int. J. Numer. Methods Fluids}, 62:\penalty0 1107--1133, 2010.

\bibitem[Zhang et~al.(2016)Zhang, Liu, Chen, and Zhong]{zhang2016}
F.~Zhang, J.~Liu, B.~Chen, and W.~Zhong.
\newblock Evaluation of rotated upwind schemes for contact discontinuity and
  strong shock.
\newblock \emph{Comput and Fluids}, 134:\penalty0 11 -- 22, 2016.

\bibitem[Peery and Imlay(1988)]{peery1988}
K.~M. Peery and S.~T. Imlay.
\newblock {Blunt-Body Flow Simulations}.
\newblock In \emph{AIAA Paper 88-2904}, 1988.

\bibitem[Lin(1995)]{lin1995}
Hong-Chia Lin.
\newblock Dissipation additions to flux-difference splitting.
\newblock \emph{Journal of Computational Physics}, 117\penalty0 (1):\penalty0
  20 -- 27, 1995.

\bibitem[Harten and Hyman(1983)]{harten_hyman1983}
A.~Harten and J.~M. Hyman.
\newblock Self adjusting grid methods for one-dimensional hyperbolic
  conservation laws.
\newblock \emph{Journal of Computational Physics}, 50\penalty0 (2):\penalty0
  235 -- 269, 1983.

\bibitem[Balsara(2010)]{balsara2010}
D.~S. Balsara.
\newblock Multidimensional hlle riemann solver: Application to euler and
  magnetohydrodynamic flows.
\newblock \emph{Journal of Computational Physics}, 229\penalty0 (6):\penalty0
  1970 -- 1993, 2010.

\bibitem[Toro(2009)]{toro2009}
E.F. Toro.
\newblock \emph{Riemann Solvers and Numerical Methods for Fluid Dynamics: A
  Practical Introduction}.
\newblock Springer Berlin Heidelberg, 2009.
\newblock ISBN 9783540498346.

\bibitem[van Leer et~al.(1987)van Leer, Thomas, Roe, and Newsome]{vanleer1987}
B.~van Leer, J.~L. Thomas, P.~L. Roe, and R.~W. Newsome.
\newblock \emph{A comparison of numerical flux formulas for the euler and
  navier-stokes equations}, pages 36--41.
\newblock American Institute of Aeronautics and Astronautics Inc, AIAA, 1987.

\bibitem[Toro and Cendon(2012)]{toro_vaz2012}
E.F. Toro and V.~Cendon.
\newblock {Flux splitting schemes for the Euler equations}.
\newblock \emph{Comput and Fluids}, 70:\penalty0 1--12, 2012.

\bibitem[Batten et~al.(1997)Batten, Clarke, Lambert, and Causon]{batten1997}
P.~Batten, N.~Clarke, C.~Lambert, and D.~M. Causon.
\newblock On the choice of wavespeeds for the hllc riemann solver.
\newblock \emph{SIAM J. Sci. Comput.}, 18\penalty0 (6):\penalty0 1553--1570,
  November 1997.

\bibitem[Mandal and Panwar(2012)]{mandal2012}
J.C. Mandal and V.~Panwar.
\newblock Robust hll-type riemann solver capable of resolving contact
  discontinuity.
\newblock \emph{Comput and Fluids}, 63:\penalty0 148 -- 164, 2012.

\bibitem[Schmidtmann and Winters(2017)]{schmidtmann2017}
B.~Schmidtmann and A.~R. Winters.
\newblock Hybrid entropy stable hll-type riemann solvers for hyperbolic
  conservation laws.
\newblock \emph{Journal of Computational Physics}, 330:\penalty0 566 -- 570,
  2017.

\bibitem[Davis(1988)]{davis1988}
S.~F. Davis.
\newblock Simplified second-order godunov-type methods.
\newblock \emph{SIAM Journal on Scientific and Statistical Computing},
  9\penalty0 (3):\penalty0 445--473, 1988.

\bibitem[Liou(2000)]{liou2000}
Meng-Sing Liou.
\newblock {Mass flux schemes and connection to shock instability}.
\newblock \emph{J. Comput. Phys.}, 160:\penalty0 623--648, 2000.

\bibitem[Zhang et~al.(2017)Zhang, Liu, Chen, and Zhong]{zhang2017}
F.~Zhang, J.~Liu, B.~Chen, and W.~Zhong.
\newblock A robust low-dissipation ausm-family scheme for numerical shock
  stability on unstructured grids.
\newblock \emph{Int. J. Numer. Methods Fluids}, 84:\penalty0 135--151, 2017.

\bibitem[Blazek(2005)]{balzek2005}
J.~Blazek.
\newblock \emph{Computational Fluid Dynamics: Principles and Applications
  (Second Edition)}.
\newblock Elsevier Science, 2005.
\newblock ISBN 9780080529677.

\bibitem[Barth and Jesperson(1989)]{barth1989}
T.J. Barth and D.C. Jesperson.
\newblock \emph{The design and application of upwind schemes on unstructured
  meshes}.
\newblock American Institute of Aeronautics and Astronautics Inc, AIAA, 1989.

\bibitem[Gottlieb et~al.(2001)Gottlieb, Shu, and Tadmor]{Gottlieb2001}
S.~Gottlieb, C.-W. Shu, and E.~Tadmor.
\newblock {Strong Stability-Preserving High-Order Time Discretization}.
\newblock \emph{SIAM Rev.}, 43\penalty0 (1):\penalty0 89--112, 2001.

\bibitem[Godunov(1959)]{godunov1959}
S.K. Godunov.
\newblock {A finite difference method for the computation of discontinuous
  solutions of the equations of fluid dynamics}.
\newblock \emph{Matematicheskii Sbornik}, 47:\penalty0 357--393, 1959.

\bibitem[Ohwada et~al.(2013)Ohwada, Adachi, Xu, and Luo]{ohwada2013}
Taku Ohwada, Ryo Adachi, Kun Xu, and Jun Luo.
\newblock {On the remedy against shock anomalies in kinetic schemes}.
\newblock \emph{Journal of Computational Physics}, 255:\penalty0 106--129,
  2013.

\bibitem[Woodward and Colella(1984)]{woodward1984}
P.~Woodward and P.~Colella.
\newblock The numerical simulation of two-dimensional fluid flow with strong
  shocks.
\newblock \emph{Journal of Computational Physics}, 54\penalty0 (1):\penalty0
  115 -- 173, 1984.

\bibitem[Wada and Liou(1997)]{wada1997}
Y.~Wada and M.-S. Liou.
\newblock An accurate and robust flux splitting scheme for shock and contact
  discontinuities.
\newblock \emph{SIAM Journal on Scientific Computing}, 18:\penalty0 633--657,
  1997.

\end{thebibliography}

\end{document}